\documentclass[letterpaper,twocolumn,10pt]{article}

\usepackage{graphicx}
\graphicspath{{figures/}}
\usepackage[latin1]{inputenc}
\usepackage{url}
\usepackage{xspace}
\usepackage{amssymb}
\usepackage{amsmath}
\usepackage{subfigure}
\usepackage{tabularx}
\usepackage[labelfont=bf, font=bf]{caption}

\makeatletter
\renewcommand*{\p@section}{\S\,}
\renewcommand*{\p@subsection}{\S\,}
\makeatother

\newcommand{\name}{SmartFlux\xspace}
\newcommand{\footurl}[1]{\footnote{\url{#1}}}
\newcommand{\myparagraph}[1]{\vspace{6pt}\noindent\textbf{#1.}}
\newcommand{\myitem}[1]{\noindent\emph{#1:}}

\begin{document}
\title{
%
}

\title{
Smart Scheduling of Continuous Data-Intensive Workflows with Machine Learning Triggered Execution
}

\author{Sérgio Esteves, Helena Galhardas, and Luís Veiga \\
INESC-ID Lisboa, Instituto Superior Técnico, Universidade de Lisboa \\
\{sergio.esteves, helena.galhardas, luis.veiga\}@tecnico.ulisboa.pt
}

\date{INESC-ID Technical Report 9/2016, Oct. 2016}





\maketitle

\begin{abstract}

%
%
%
%
To extract value from evergrowing volumes of data, coming from a number of different sources, and to drive decision making, organizations frequently resort to the composition of data processing workflows, since they are expressive, flexible, and scalable. The typical workflow model enforces strict temporal synchronization across processing steps without accounting the actual effect of intermediate computations on the final workflow output. However, this is not the most desirable behavior in a multitude of scenarios. We identify a class of applications for continuous data processing where workflow output changes slowly and without great significance in a short-to-medium time window, thus wasting compute resources and energy with current approaches.
\\\indent To overcome such inefficiency, we introduce a novel workflow model, for continuous and data-intensive processing, capable of relaxing triggering semantics according to the impact input data is assessed to have on changing the workflow output. To assess this impact, learn the correlation between input and output variation, and guarantee correctness within a given tolerated error constant, we rely on Machine Learning. The functionality of this workflow model is implemented in \name, a middleware framework which can be effortlessly integrated with existing workflow managers. Experimental results indicate we are able to save a significant amount of resources while not deviating the workflow output beyond a small error constant with high confidence level.




\end{abstract}

\section{Introduction}
\label{sect:introduction}
Current trends are being characterized by an ever-growing volume of data flowing over the globe throughout wide-scale networks. To face this, new distributed and high-scalable infrastructures are required to manage and process data efficiently. The trend has been to move towards infrastructures enabling workflow composition, denominated Workflow Management Systems (WMSs), since they enable better expressiveness, flexibility, and maintainability when compared with lower-level code (e.g., Java map-reduce code).

A workflow is usually modeled as a Directed Acyclic Graph (DAG) to express the dependencies and relations between computation and data. 
The workflow paradigm has been extensively used in a number of different settings (e.g., eScience, engineering, industrial), encompassing activities as diverse as web crawling, data mining, protein folding, sky surveys, forecasting, RNA-sequencing, or seismology~\cite{Ahrens:2011:DSU:2086320.2086406,1538-3881-120-3-1579,citeulike:3299862,Deelman:2006:MLW:1191828.1192568}.

A WMS is different from a Stream Processing System (SPS). In a WMS, computations are triggered by time and data availability (discrete events in time), and are not based on sliding time windows (like in SPS). WMS accumulate, persist and communicate larger quantities of data across processing steps, whereas SPS keep most of the data in volatile memory. Continuous processing in the context of this work means that the same aggregated computation (workflow) is executed multiple times over ("non-contiguous") time, and does not necessarily mean that workflows are uninterruptedly receiving new input data (like in SPS).

Traditionally, WMSs enforce strict temporal synchronization throughout the various dependencies of processing steps (i.e., following the Synchronous Data-Flow (SDF) computing model \cite{ludaescher09:_scien_proces_autom_and_workf_manag}). That is, a step is immediately triggered for execution as soon as all its predecessor steps have finished their execution. Should the temporal logic be relaxed, for example, to respond to application requirements of latency or prioritization, programmers have no other choice than to explicitly program non-synchronous behavior. This ad-hoc programming increases the complexity of the application and the chance of error occurrence.

In addition, typical WMSs do not take into account the volume of data arriving at each processing step and its actual impact on changing the final workflow output (i.e., the output produced by processing steps that do not have any successor steps). We argue that such an assessment should be used to control the workflow execution and drive the triggering of steps towards meaningful results. This issue is even more important in workflows for data-intensive and continuous processing where many resources can be purposelessly wasted if new input and intermediate datasets do not cause significant changes on workflow output across complete executions.

In fact, fully executing a processing step every time a small fragment of data is received can have a great impact on performance and machine load, without actually changing substantially the workflow output (or foremost, its significance to the problem being addressed); as opposed to executing it only when a certain substantial, relevant (w.r.t. application semantics) quantity of new data is available.

\begin{figure}
  \centering
  \includegraphics[width=0.9\columnwidth]{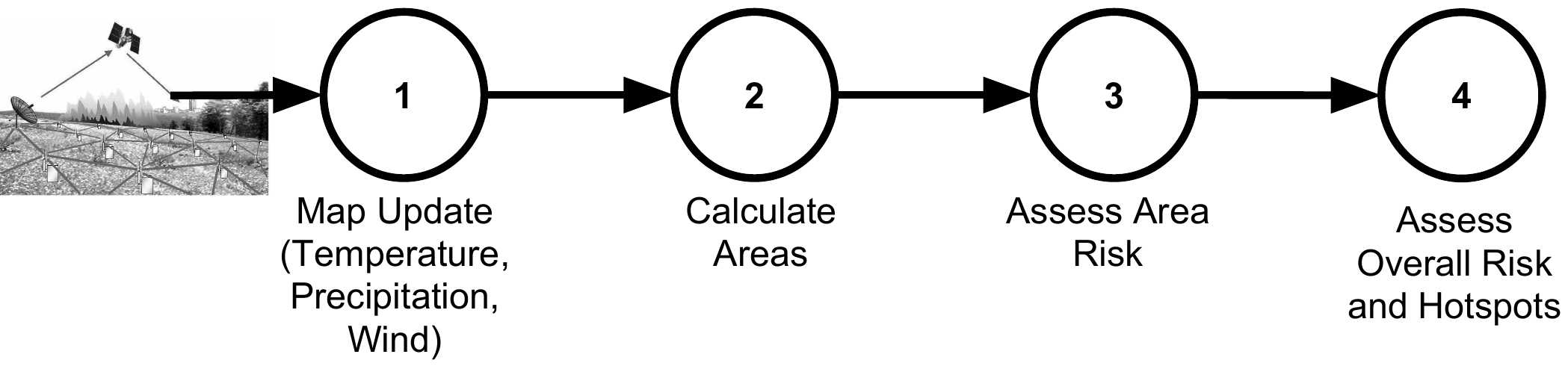}
  \caption{Motivational example: fire risk assessment}
  \label{fig:simple-fire-forecast}
\end{figure}

Further, there is a class of workflow applications for continuous data processing where the output of final processing steps does not change significantly in a short-to-medium time window. As a motivational example, consider the case of assessing the fire risk in a given forest through a sensor network that captures temperature, precipitation and wind (a more complete description is given in cc). Figure~\ref{fig:simple-fire-forecast} depicts a workflow that, periodically (e.g., every 5 seconds, every half an hour), receives data from the sensors and executes the following processing steps: 1) updates an internal representation of the forest map; 2) calculates areas by dividing the map; 3) assesses the fire risk in each area; and 4) assesses the overall risk and contiguous risky areas (hotspots).

The temperature, precipitation and wind measures will probably not change every half an hour, or at least not significantly to pose a risk. Changes in the sensor readings will cause increasingly smaller changes in the data as we go through the steps of the workflow; e.g., the temperature of an area (one piece of data generated by step 2), which results from the aggregation (average) of the temperature of its composing sensors, will only change if a large fraction of sensor readings change (more than one piece of data in step 1 should change). Likewise, the risk of an area (one piece of data generated by step 3), which consists of the classification of different temperature ranges into different risk levels, will probably only change after an area has been updated several times (many more than one piece of data in the output of step 2). As a result, the output of the workflow, generated by step 4, will remain almost unchanged during most of the time. Only output variations higher than a certain threshold (i.e., decision-making boundary) will be deemed as significant. Therefore, we consider that a substantial amount of resources is wasted in re-executing the entire workflow.

Other examples, that fall in this same application class, include: measuring the impact of social business~\cite{sbusinessindex}, 
detecting gravitational-waves~\cite{springerlink:10.1007/978-1-84628-757-2_4}, weather forecasting~\cite{springerlink:10.1007/s12145-008-0010-7}, predicting earthquakes~\cite{Deelman:2006:MLW:1191828.1192568}, among others. Even for those applications where the outcome changes more frequently, such as a web crawler, the impact of the updated results may become only relevant when the differences from the previous crawls accumulate significantly (e.g., relevant change in word counts, page ranking or the number of reverse links).

In this paper, we address the problem of providing asynchrony in workflows resorting to the notion of Quality-of-Data. We define Quality-of-Data (QoD),\footnote{Quality-of-Data is akin to Quality-of-Service, and should not be confused with issues such as internal data correctness, semantic coherence, data adherence to real-life
sources, or data appropriateness for managerial and business decisions.} in this context, as the entirety of features or characteristics that data must have towards its ability to satisfy the purpose of changing the workflow output significantly w.r.t. the application specific semantics (following the principle described in~\cite{juran1999juran}). With this notion, which is akin to Quality-of-Service, we are thus able to assign different priorities to different data sets, users or workflows, or to guarantee a certain level of performance in a workflow. These performance guarantees can be enforced, for example, based on the size and magnitude of new updates. Particularly, we enforce QoD guarantees based on metrics collected on distributed Key-Value data stores (e.g., Cassandra~\cite{Lakshman:2009:CSS:1582716.1582722}, HBase~\cite{hbase2011george}).


We introduce a novel workflow model, for data-intensive and continuous processing, that is capable of intelligently guiding the triggering of processing steps, according to patterns observed in the flow of data, towards a meaningful and significant output, while respecting QoD constraints. To assess how different input data patterns affect the workflow output, we resort to Machine Learning with Random Forests~\cite{Breiman:2001:RF:570181.570182}, which is the classification algorithm that yielded better performance in general comparing to others (cf. \ref{sect:learning}). Specifically, we learn statistical behaviors of workflows by correlating input variation with output generated deviation, arising from skipping the execution of processing steps.

To the best of our knowledge, this is the first model, in general DAG processing, that can skip computations based on the predicted impact they have on changing the workflow output. Specifically, we trade-off resource savings with result accuracy by allowing (small) errors to exist in the output. When the error is above a given threshold (possibly representing a decision-making boundary), it means that the output is significant and computations should be therefore executed.

Note that we do not discard any data, like it happens in load shedding~\cite{Tatbul:2007:SFE:1325851.1325873}. In load shedding, a fraction of the input data is shed to alleviate overloaded servers and preserve low latency for query results. Contrarily to discarding data, we accumulate it up to the point where it causes significant changes on the output of the workflow. The observed errors occur not because we are making computations with incomplete data, but because we are not performing the computations and generating new output (i.e., errors come from stale data in the output).

There has also been a recent effort to enable approximate processing in data processing systems (e.g., MapReduce, Stream Processing) in order to reduce latency (and possibly resource usage). However, these systems usually only target specific aggregation operators (e.g., sum, count) in structured languages~\cite{Goiri:2015:ABA:2694344.2694351,agarwal:2013:bqb:2465351.2465355,Agarwal:2014:KYW:2588555.2593667}. In our work, we provide approximate results for general-purpose computations; i.e., we are agnostic to the code that is running on each processing step and solely observe the data that is inputted and its effect on modifying the output. Effectively learning the correlation between input and output enables us to bound the error and give guarantees about the correctness of the results.

As a proof of concept, we developed \name, a middleware framework that enforces our asynchronous model and can be integrated with existing WMSs. In this work we integrate it with a widely-deployed WMS, Apache Oozie~\cite{islam:2012:ots:2443416.2443420}. Our experimental results indicate that, with \name, we are able to deliver high resource efficiency. Specifically, we are able to save a significant amount of resources while not deviating the workflow output beyond a small error constant with a high confidence level: up to 30\% less executions while enforcing a QoD (an error bound) as low as 5\% with a confidence over 95\%.

The main contributions of this paper are: i) a novel workflow model that enables triggering asynchrony across processing steps; and ii) a framework (\name) that uses Random Forests to guide workflow execution towards meaningful results (w.r.t. application semantics) in a resource-efficient manner.

The remainder of this paper is structured as follows. \ref{sect:model} details our workflow model. \ref{sect:learning} describes our learning approach to bound the output error. \ref{sect:architecture} presents the design and architecture of \name and \ref{sect:evaluation} its experimental evaluation. Related work follows in \ref{sect:related-work} and \ref{sect:conclusion} concludes the paper.

\section{Abstract Workflow Model}
\label{sect:model}
In this section we describe our workflow model that enables temporal asynchrony, by allowing flexible control of the triggering of processing steps, based on the predicted impact that observed data patterns in the input will have on the workflow output. Our model is specifically designed for continuous processing and data-intensive scenarios where long-lasting workflow applications are regularly fed with new raw data from a given source (e.g., network of sensors, Internet, social network, radio telescopes). We refer to each time a workflow is fed with new data as a \textit{wave}.

Our workflow model inherits from and extends the traditional workflow model~\cite{workflow-survey-2005} where strict temporal synchronization is enforced. As we are targeting data-intensive applications, our focus in this work regarding data communication is put on distributed Key-Value stores, since they can achieve better scalability, locality-awareness, and flexibility than just using a file system. Our model could be adapted to operate with (unstructured) files, but such adaptation is out of the scope of this paper. Plus, we perform specific analysis and processing that is oriented by Key-Value abstractions (e.g., to compute metrics regarding new data updates). Hence, we use this type of storage as an advantage, and it can fit many, if not most, large scale data processing scenarios.

In distributed Key-Value stores, like the columnar-oriented HBase and Cassandra, data containers may consist of keyspaces, tables, columns (including hierarchical columns), rows, or any combination of these, and it is usually trivial, through simple get-put interfaces, to capture the scope of update operations in terms of affected containers; i.e., there is no need to deal with wider-scope queries possibly containing complex aggregate and join operations. We define an \emph{element} in a data container as a (multi-dimensional) Key-Value pair.

The main feature that differentiates our model from the other typical DAG workflows is its triggering semantics: a processing step A, in a workflow D, is not necessarily triggered for execution immediately, when all its predecessors A' ($A' \prec_D A$) have finished their execution. Instead, A should only be triggered as soon as all predecessor steps A' have completed at least one execution and have, also, carried out a sufficient (or significant) level of changes on the underlying KV store that comply with certain QoD requirements. This way, a processing step can be re-executed several times without necessarily triggering the execution of successor nodes; i.e., the triggering of steps is guided by the rate of data changes, and not exclusively by the end of a single execution of predecessor nodes, as it usually happens in the regular workflow model. This enhanced semantics-guided incremental behavior can improve expressiveness, e.g. w.r.t. Percolator~\cite{Peng:2010:LIP:1924943.1924961}.

The QoD, that a processing step needs to comply with, corresponds to the impact on its input (which comes from the generated output of predecessor steps) that makes its output reach a maximum defined tolerated error. Hence, the target impact on input corresponds to the input necessary, in terms of quantity and quality (or significance), for reaching a threshold that specifies the maximum deviation of the output tolerated for that step. This output deviation in a step can be seen as an error introduced by delaying and skipping its (re-)execution, as opposed to the synchronous model. In return, delaying and skipping execution save resources from being wastefully engaged. Following, we describe the metrics to calculate the input impact and the output error.

\myparagraph{Input Impact}
The input impact of a processing step is a metric that captures the amount and magnitude of changes performed on its associated data container (e.g., a column or a set of columns) in relation to a previous state. Every time new data updates are performed on a data container that holds the input of a processing step, the input impact is calculated based on the new updated data and their previous versions. The previous versions correspond either to the state of the data on the previous wave, or the state of the data on the wave where the latest execution of the step occurred. The former implies that the input impact is accumulated with the impact measured for previous waves that occurred after the execution of the associated step; while the later allows computations to cancel each other out: if we get the value $x_i$ on wave $w$ equal to $x_i'$ on wave $y$, and regardless of the number of waves occurred between $y$ and $w$ without triggering the associated step, the error comes as zero. Further, since steps are potentially not all executed at the same wave, the input impact of a step is only calculated when its predecessors have generated output, which will possibly not happen in every wave.

We provide an API through which users can define their own functions to capture the impact of changes in a data container (elaborated in \ref{sect:api}). Our API comes with two base implementations that represent two different yet generic functions that can serve well a wide set of scenarios according to our experiments (input impact is denoted as $\iota$). They are described in the following equations.
%
%

\hspace{-25pt}\begin{tabularx}{1.1\columnwidth}{@{}XXX@{}}
\begin{equation}
\iota=\sum\limits_{i=1}^m |x_i - x_i'| \times m
\label{eq:input-impact-2}
\end{equation} &
\begin{equation}
\iota=\frac{\sum_{i=1}^m |x_i - x_i'| \times m}{\sum_{i=1}^m max(x_i, x_i')\times n}
\label{eq:input-impact-3}
\end{equation}
\end{tabularx}

In the equations above, $x_i$ is the updated state of the ith element and $x_i'$ its latest state, $m$ and $n$ are the number of modified elements and the total number of elements in the associated data container respectively, and $max$ is the function that returns the maximum between two numbers. If a new element is inserted, its latest state $x_i'$ is zero (which increases the impact).

Equation~\ref{eq:input-impact-2} captures the differences in magnitude between the updated and latest snapped state of elements, multiplied by the number of modified elements. Equation~\ref{eq:input-impact-3} divides the result of Equation~\ref{eq:input-impact-2} by the maximum between the updated and latest state of the elements, multiplied by the total number of elements in the data container. Hence, it captures the relative impact over a previous state, returning a value between $0$ (no changes) and max out at $1$ (difference introduced by new data with higher or equal magnitude of the previous state).


Further, if a processing step receives input from more than one predecessor step, then we calculate the input impact produced by each predecessor step and combine them through the geometric mean (albeit other aggregations can be applied).


\myparagraph{Output Error}
The output error of a processing step is a metric that attempts to measure the error penalty (or impact) of postponing its executions. Each time a step is not executed at a given wave of data, it incurs a certain error that can be seen as the cost of the changes that were missed in the corresponding data container. Hence, if a step is always executed at each wave of data the error is zero. Like the input impact, the output error can be cumulative or not depending on whether error cancellation is allowed for an application.

Users also have the flexibility of providing their own implementations of our API to compute the output error (cf. \ref{sect:api}). As base implementations, we offer the following generic functions to calculate the output error (denoted by $\varepsilon$).


\hspace{-25pt}\begin{tabularx}{1.1\columnwidth}{@{}XXX@{}}
\begin{equation}
  \varepsilon=\frac{\sum_{i=1}^m |x_i - x_i'| \times m}{\sum_{i=1}^n x_i' \times n}
  \label{eq:error-2}
\end{equation} &
\begin{equation}
\varepsilon=\sqrt{\frac{\sum_{i=1}^m (x_i - x_i')^2}{m}}
  \label{eq:error-1}
\end{equation}
\end{tabularx}

In the equations above, $x_i$ is the updated state of the ith element and $x_i'$ its latest state, $m$ and $n$ are the number of modified elements and the total number of elements in the associated data container respectively.

Equation~\ref{eq:error-2} 
captures the relative impact of the difference to the correct state, value between 0 (no error) and 1 (new data has higher or equal magnitude of the previous state). Equation~\ref{eq:error-1} corresponds to the frequently used Root-Mean-Square Error (RMSE), which captures the deviation between the updated and previous states of elements, thereby attenuating the impact of small differences and penalizing larger differences more. It is up to the user to decide which function works better for a particular problem.


For each considered processing step, the output error $\varepsilon$ must be bounded in order to ensure, with  a given confidence, an acceptable level of correctness and usefulness in the result of the workflow (i.e., results should not be significantly deviated in relation to the normal execution of the synchronous model). During a period of normal execution, we learn statistical behavior of the errors for different impacts of input given to processing steps (elaborated in \ref{sect:learning}).

\myparagraph{Generality of the Model}
\label{sect:generality}
Our model is suitable for applications that exhibit similar input patterns over a period of time (i.e., no random or uncorrelated input/output over time). This class of applications is commonplace in the domain of continuous workflow processing. As long as there is a correlation between input and output, our system is able to accurately predict, with a high confidence interval, when and which steps should be skipped or executed.

Hence, this is the central premise for our system to work. Following, we briefly describe that there is an intuitive relation between input and output for three pipeline/workflow applications (due to space constraints we abstract from the details of processing steps that perform the computations).

\myitem{PageRank} Processes the content of crawled documents and builds an histogram with the differences against previous states of links. It is only worthy to process the new crawled documents if the differences in the link counts are sufficient to significantly change the page rank of documents (according to decision makers).

\myitem{LIGO \cite{springerlink:10.1007/978-1-84628-757-2_4}} It detects gravitational waves that are linked to the occurrence of events in the universe. The output, regarding the detection of events (like exploding stars), is strongly associated with the input which corresponds to laser data waveforms. It is only worthy to explore the input data if the simple characterization of waveforms can lead to a true inspiral (event).

\myitem{CyberShake \cite{Deelman:2006:MLW:1191828.1192568}} It performs seismic hazard estimation for a given site. The input corresponds to rupture descriptions and the output is an hazard map. It is only worthy to recompute parts of the map if the new probability variations of ruptures are impactful against a previous state.

In these, as well as in a great part of applications for continuous and incremental processing, there is generally an association between input and output. The correct characterization of that input can allow us to predict the significance on the output.
%

\myparagraph{Prototypical Scenario}
\label{sect:prototypical-scenario}
\begin{figure}
	\includegraphics[width=\columnwidth]{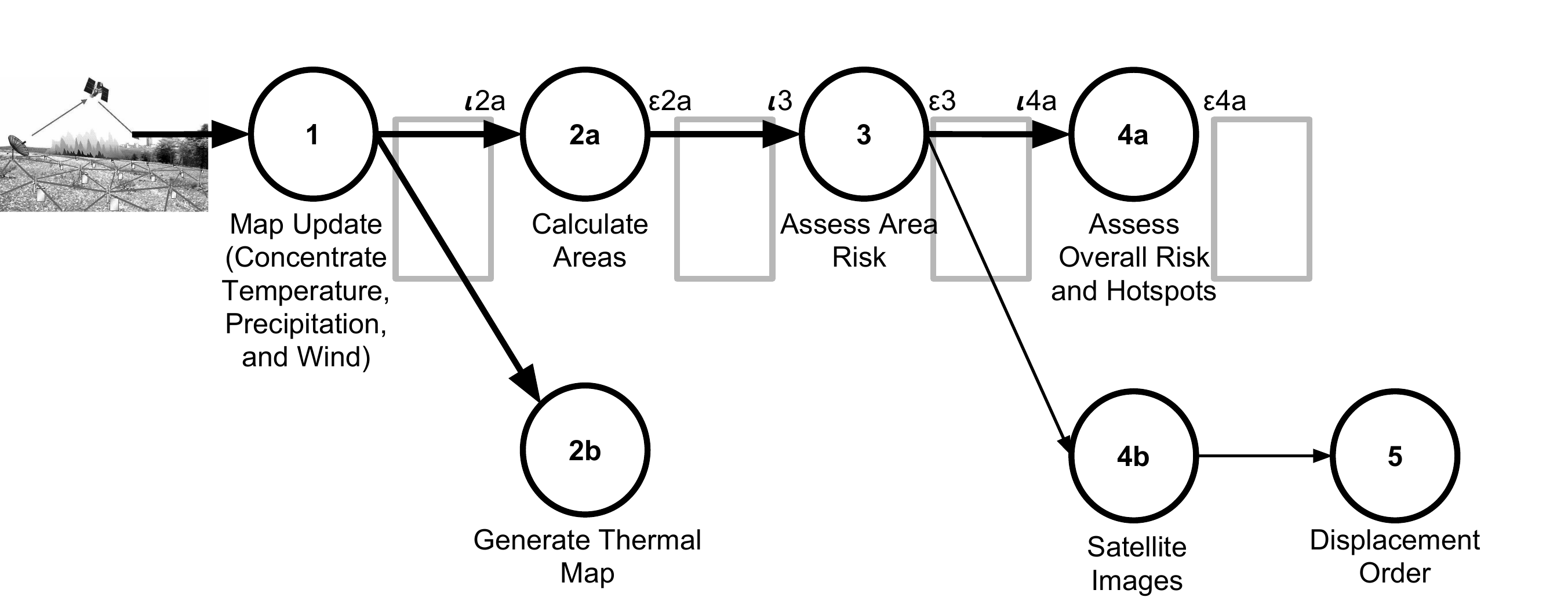}
	\caption{Use-case scenario of continuous and incremental workflow processing: fire risk assessment. The rectangles in grey represent columnar-like data containers.}
	\label{fig:fire-forecast-workflow}
\end{figure}
\begin{figure}
    \centering
    \includegraphics[width=0.7\columnwidth]{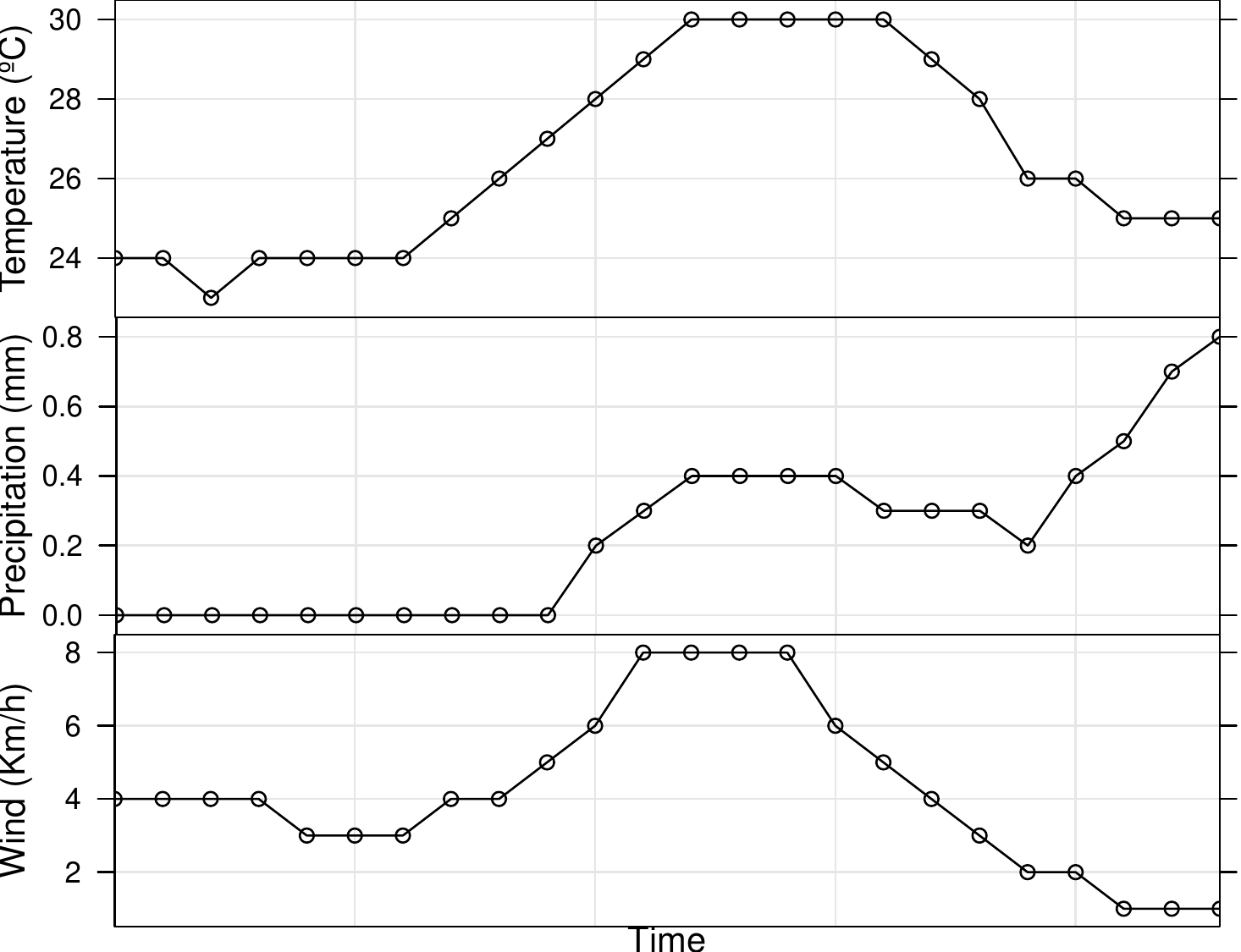}
    \caption{Temperature, precipitation and wind evolution hour by hour for a day in the Amazon rainforest}
    \label{fig:weather-evolution}
\end{figure}
Figure~\ref{fig:fire-forecast-workflow} illustrates a workflow that assesses the fire risk for a given forest region based on a network of sensors equally distributed. For a normal day in the Amazon rainforest, for instance, we can see in Figure~\ref{fig:weather-evolution} that temperature, precipitation, and wind, vary progressively over 24 hours (this also holds, even more so, when we assume higher frequency of sensor readings, e.g., every second) and without major steep slopes. Such characteristics make this scenario propitious for resource reasoning and savings.

The first processing step (in Figure~\ref{fig:fire-forecast-workflow}) receives data from sensors every time interval (temperature, precipitation, wind), aggregates it through some function, and stores the result for each sensor. Since this is the first step that updates a data container, it must always be executed at every wave (i.e., it is not possible to maintain sensory data across waves without the execution of this step). \textbf{Step 2a} divides the forest into smaller areas and combines the measures of all sensors in each area. This step is only executed when $\iota_{2a}$ is sufficient to cause $\varepsilon_{2a}$ to reach $max(\varepsilon_{2a})$, which is the (user-defined) maximum tolerated error. \textbf{Step 2b} generates a thermal graphical map for some monitoring station.

\textbf{Step 3} assesses the fire risk of each area by comparing the values calculated in \textbf{step 2a} with some threshold. This step is only triggered when significant measurement differences in some areas are perceived or when a sufficient number of areas is updated by \textbf{step 2a}.

\textbf{Step 4a}, which is the workflow output, assesses the overall fire risk and identifies groups of areas with higher risk in the forest. This step is expected to have its output changed slowly over time and $max(\varepsilon_{4a})$ should be set to a value such that the difference in the overall fire risk across waves is significant to decision makers. \textbf{Step 4b} gathers satellite images in case areas identified in \textbf{step 3} are with very high temperature levels (on fire); and \textbf{step 5} issues a displacement order to a fire department in case the fire is confirmed through the analysis of satellite images. These two last steps are critical for fire detection and therefore they do not tolerate error.

To estimate and correlate error with input impact, we learn the statistical behavior of the workflow with Machine Learning by executing the workflow synchronously for a restricted period of time, as we elaborate next.

\section{Learning Approach}
\label{sect:learning}
This section introduces our learning approach to bound the output error, arising from the delayed execution of processing steps, and to provide guarantees about the maximum deviation of workflow outputs. Specifically, we make use of Machine Learning classification techniques to predict how input data affects the output of processing steps.

In fact, our learning approach is based on predictions that are not perfect (albeit we can get very close approximations in general), and therefore the guarantees we refer in this paper are \emph{probabilistic} guarantees; i.e., we are able to ensure that error bounds are respected \emph{within} a confidence interval (these are the same kind of guarantees offered by other systems such as~\cite{Theobald:2004:TQE:1316689.1316746,agarwal:2013:bqb:2465351.2465355}). This confidence interval is expected to be high ($> 90\%$) as long as our central premise holds; i.e., that there is a correlation between input and output (cf., \ref{sect:generality}). This premise is verified during a test phase (elaborated later on in this section).

\myparagraph{Algorithm Selection} 
To select a good Machine Learning classification algorithm for our problem, we performed several experiments using the applications described in \ref{sect:evaluation}. Through the ROC area, a metric to assess the performance of a classifier, we compared the following widely-deployed algorithms: Bayes Network, J48 tree, Logistic, Neuronal Network, Random Forest, and Support Vector Machine. Random Forest (RF)~\cite{Breiman:2001:RF:570181.570182} and Support Vector Machine (SVM)~\cite{hearst1998support} yielded better ROC areas on average for all the experiments: 0.86 and 0.82 respectively (values approaching 1 mean optimal classifier and 0.5 being comparable to random guessing). However, since SVM requires more parameterization (e.g., selecting a proper kernel to capture linear or non-linear data correlations, or using cost matrices to weight unbalanced datasets)~\cite{steinwart2008support}, and default parameterization in RF often performs well~
\cite{Breiman:2001:RF:570181.570182}, we decided to adopt RF as our default learning approach to all experiments (albeit the algorithm can be easily switched).

\myparagraph{Classification} 
Generally, classification algorithms try to estimate a function $h(x)$ that, given a set with \emph{N}-dimensional input data, predicts which of two possible classes form the output ($f:\mathbb{R}^N\rightarrow\lbrace\pm 1\rbrace$). The estimation of this function, which corresponds to the construction of a model, is based on a supplied set of training examples encompassing tuples with known correct values of input and corresponding output (i.e., supervised learning). The obtained classifier is then able to assign new unseen examples to one class or another.


In our particular problem, we need to predict which steps generate an error exceeding their corresponding maximum bounds ($max_\varepsilon$) for a given input. Hence, the output of the classifier is no longer a single binary value, but a set of values representing the configuration of steps that should be executed or not for each wave of data (i.e., multi-label classification~\cite{tsoumakas2007multi}). For example, the matrices below represent, for 5 waves, a pipeline with 3 steps querying the classifier by sending the input impact $\iota$ calculated for each step ($X$), and receiving in return the sequence of steps that should be executed or not ($Y$).


$h(\underbrace{\begin{bmatrix}
 694.86 & 601.6 & 498.3 \\
 191.24 & 886.1 & 498.3 \\
 278.13 & 1071.4 & 498.3 \\
 433.78 & 233.78 & 664.24 \\
 551.53 & 523.8 & 956.52
\end{bmatrix}}_{X})=
\underbrace{
\begin{bmatrix}
  1 & 0 & 0 \\
  0 & 0 & 0 \\
  0 & 1 & 0 \\
  0 & 0 & 0 \\
  0 & 0 & 1
\end{bmatrix}}_{Y}$

To learn the correlations between input impact and respective incurred error, it is necessary to train the RF classifier and construct a model. After, in a test phase, the quality of the trained classifier is assessed and, if the accuracy is not satisfactory, more training may be required. These two sequential phases, training and test, can be performed either regularly from time to time or on-demand (useful if data patterns start to change suddenly). Further, these phases take place while the workflow is running and producing results with real datasets, hence making this an online process.

\myparagraph{Training Phase}
Unless a training set is given beforehand, a training phase starts taking place when the workflow is executed for the first time. During this phase, all processing steps of the workflow are executed synchronously (without any QoD enforcement) for a given configurable number of waves or time frame. At each wave, the input impact $\iota$ and corresponding (simulated) output error $\varepsilon$ are calculated for each step, and a tuple containing $\iota$ and a binary value, indicating whether the $\max_\varepsilon$ of that step is reached, is appended to a log (corresponding to our training-set).

\myparagraph{Test Phase}
In the test phase, we assess the quality of the trained model thereby measuring namely: i) accuracy, proportion of instances correctly classified; ii) precision, number of instances that are truly of a class divided by the total instances classified as that class; and iii) recall, number of instances classified as a given class divided by the actual total of that class. We perform a 10-fold cross-validation on the training-set.

High values of recall mean that we are avoiding the existence of false negatives; i.e., the percentage of times the model estimated incorrectly that the error was below $max_\varepsilon$. Hence, a good recall is necessary to ensure that the error stays within $max_\varepsilon$. As for precision, high values mean that we are avoiding to estimate incorrectly the error as being above $max_\varepsilon$, which is necessary to mitigate resource waste.

The algorithm (RF) can be adjusted to favor results on a given metric (e.g., recall), and to specify whether it is more important to comply with error bounds or save resources. If results are not satisfactory, w.r.t. defined thresholds, a training phase takes place again and more instances are collected. Otherwise, it means that we are able to provide probabilistic guarantees regarding error compliance. As there is a correlation between input and output, it is always possible to get a satisfactory result (e.g., over 90\% accuracy) with more training.

\myparagraph{Application Phase}
After a sufficiently accurate model is built, the application phase takes place and the workflow starts running asynchronously. For all steps at each wave, the input impact $\iota$ is calculated and fed to the classifier, which in return indicates which steps should be executed.

\section{\name Design and Implementation}
\label{sect:architecture}
\name is a middleware framework that provides functionality conforming to the workflow model described in the previous sections. It couples a WMS with a data storage system by monitoring data transfers and controlling the triggering of processing steps. With this coupling, \name enables the deployment of quality-driven workflow applications, where processing steps are triggered based on the  impact their computations are predicted to have in the final workflow output.

\begin{figure}
  \centering
  \includegraphics[width=0.6\columnwidth]{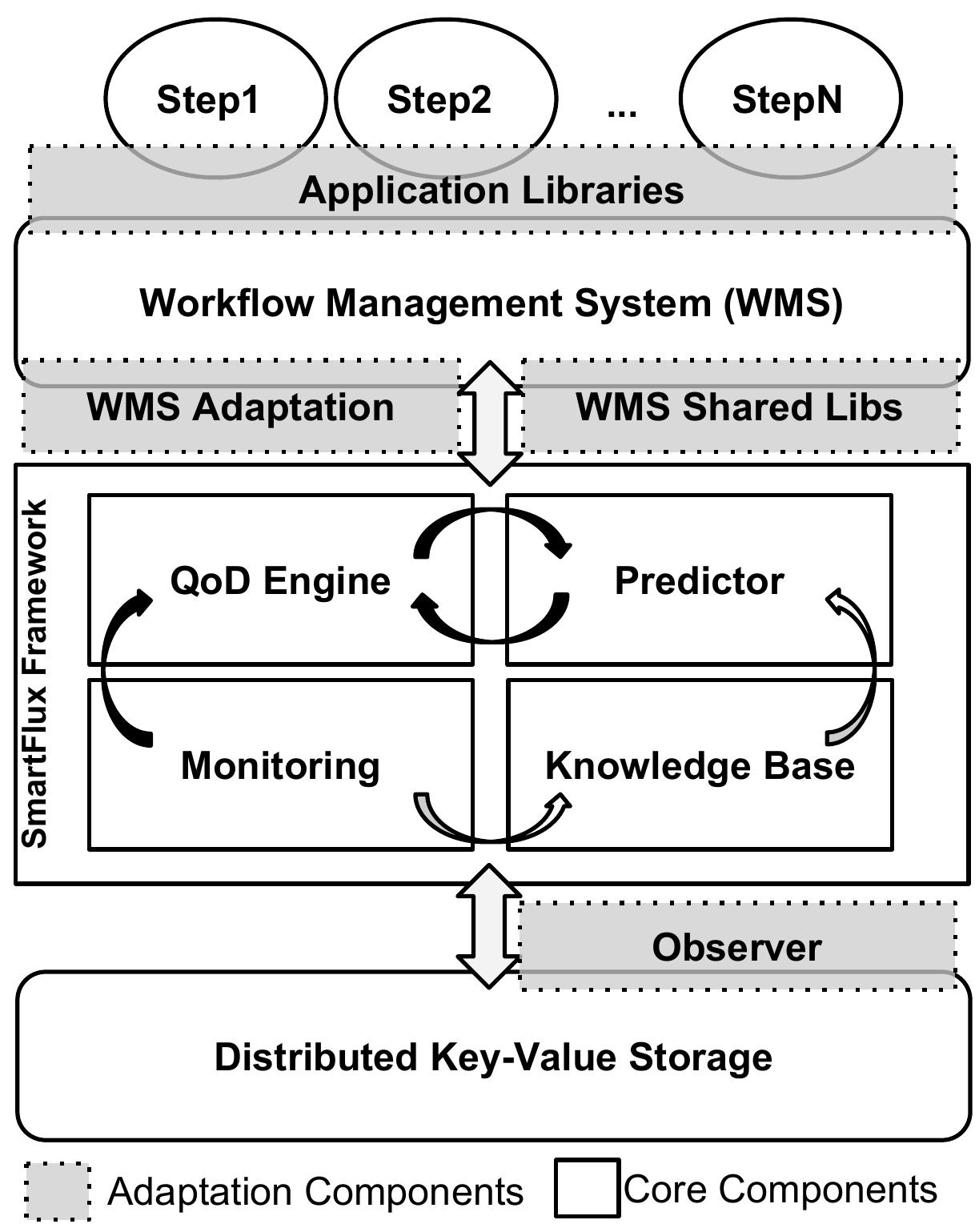}
  \caption{\name Framework Architecture}
  \label{fig:middleware}
\end{figure}

Figure~\ref{fig:middleware} illustrates the architecture of the \name middleware framework, which operates between a WMS and a (distributed) Key-Value storage system. Processing steps run atop the workflow manager and they must share data through the underlying storage system. These steps may consist of Java applications, scripts expressed through high-level languages for data analysis (e.g., Apache Pig~\cite{Olston:2008:PLN:1376616.1376726}), Map-Reduce jobs, as well as other off-the-shelf solutions.

\name can work with either its own provided simplistic WMS or existing open-source WMS. In this work we focus on using existing WMS, to assess in what extent it requires changing its implementation and triggering mechanisms. To connect our framework with a WMS, an adaptation component, \emph{WMS Adaptation} (colored in grey), needs to be provided with a specific API so that \name can issue triggering notifications and receive state information, thereby orchestrating the execution of the processing steps of a workflow.

Since \name needs to be aware of the updates the processing steps apply to the data store, we provide three options (colored in grey): i) Application Libraries, ii) WMS shared libraries, and iii) Observer. The \emph{Application Libraries} component corresponds to adapted driver libraries, used by processing steps to interact directly with the data store via their client APIs. Although applications might need to be slightly modified (e.g., changing package names in the imports of Java classes), we provide tools to completely automatize this process.

At the WMS level, \emph{WMS Shared Libraries} represent adapted shared libraries that are used by processing steps to interact with the data store through the WMS (e.g., pig scripts or any other high-level language that must be interpreted/compiled by the WMS). Finally, at a lower level, the \emph{Observer} component corresponds to custom code that is triggered and executed at the data store level upon client requests (e.g., co-processors in HBase or triggers in Cassandra). These two last options provide transparency to executing steps and avoid changes in the application code.

Next, we describe the responsibilities and purpose of each of the core components that compose the \name framework (in white). \\

\myitem{Monitoring} It analyzes, through the adaptation components, all requests directed to the Key-Value storage. This involves identifying all affected data containers and calculating the corresponding input impact and, during the training phase, also the error. Note that the simplicity of get-put interfaces work in our favor to this process. Afterwards, the calculated values are sent to the QoD Engine.

\myitem{QoD Engine} It maintains the current state of control data (input impact, error) along with workflow specification and meta-data defined by the user, such as the error bounds for each step (or data container). Based on this data, and after querying the Predictor, it evaluates and decides when and which steps should be triggered for execution during the application phase.

\myitem{Knowledge Base} It maintains data collected through the Monitoring component during the training phase: input impact and a binary value indicating whether $\varepsilon> max_\varepsilon$ for each considered step. This data forms the training-set that is used by the Predictor to build a classification model.

\myitem{Predictor} It answers to QoD Engine queries thereby predicting which error bounds are exceeded given the input impact of considered steps. For that, it uses a classifier (RF by default) with a trained model.

\myparagraph{General Work Flow}
We consider two different operating modes: i) training mode; and ii) execution mode. In the training mode, a workflow is executed synchronously and we collect metrics about the input impact and output deviation for each processing step that tolerates error. After a predetermined number of waves, a classification model is built with the previous collected data.

The training mode is represented by the white curved arrows in Figure~\ref{fig:middleware}: the Monitoring component, that gets data from the adaptation components, feeds the Knowledge Base with statistical information about the data updated in the data store; then, the Predictor component builds a classification model based on the data sets with the metrics contained in the Knowledge Base (input impact, error).

The execution mode is represented by the dark curved arrows: the Monitoring component collects statistical information from data store requests, and sends to the QoD Engine computed input impact metrics for each wave of data; after, the QoD Engine queries the Predictor with input impact data and gets in return the configuration of processing steps that should be executed in that wave (i.e., steps whose error is predicted to surpass maximum defined tolerated errors).

\myparagraph{Adopted Technology and Integration}
We integrated our framework with a widely deployed WMS, Oozie~\cite{islam:2012:ots:2443416.2443420}. In effect, we adapted Oozie by replacing the time-based and data detection triggering mechanisms, with a notification scheme that is interfaced with the \name framework process through Java RMI. Generally, Oozie only has to notify when a step finishes its execution, and \name only has to signal the triggering of a certain step; naturally, these notifications share the same processing step identifiers. The QoD error bounds are specified along with standard Oozie XML schemas (version 0.2), and given to \name with an associated workflow description. Specifically, we changed the XSD to accept a new element inside the element action (i.e., processing step) which specifies the data containers associated with steps (table, column, row, or group of any of these) and their corresponding error bounds, which are values from 0 to 1.

As our underlying distributed Key-Value storage, we adopted HBase~\cite{hbase2011george}, the open-source Java clone of BigTable~\cite{Chang:2006:BDS:1267308.1267323}. This column-oriented data store is a sparse, multi-dimensional sorted map, indexed by row, column, and timestamp; the mapped values are simply an uninterpreted array of bytes. Due to its complexity, we decided to intercept data store updates by adapting the HBase client libraries. To this end, we extended the implementation of some library classes while maintaining their original API; namely, sending the data containers and respective data to \name inside writing methods (e.g., put, delete). 
Since our API is the same, only import declarations need to be modified to \name packages in the application code.

Regarding our Machine Learning implementation, we adopted MEKA~\cite{DBLP:journals/jmlr/ReadBHP11}, a multi-label classification library in Java based on the well known WEKA~\cite{Hall:2009:WDM:1656274.1656278} Toolkit.

\myparagraph{Input Impact and Output Error API}
\label{sect:api}
We provide an API through which users can implement custom functions to capture the input impact and corresponding output error. This API comprises 3 Java method signatures that need to be implemented: \emph{processElement}, \emph{aggregate}, and \emph{compute}. \emph{processElement} is called on every element in a data container, receiving as arguments the current and previous values of the element, and returning a numeric value or tuple (e.g., it can return the difference in absolute value between the current and previous value of element). \emph{aggregate} works like a reducer, thereby aggregating pairs of values returned by \emph{processElement} (it starts reducing while \emph{processElement} is still being called to avoid accumulating a large number of values in memory), and returns a numeric value or tuple. Finally, \emph{compute} is called with the values of the last \emph{aggregate} and returns a numeric value corresponding to the overall input impact or output error of a step. To make this process easier for non-expert users, we have plans to offer an expressive high-level DSL language for the future.


%
%
%
%
%
%
%

\section{Experimental Evaluation}
\label{sect:evaluation}
In this section we present the experimental evaluation of our workflow model with the \name framework. First, we show the patterns correlating input impact with error and why Machine Learning is needed. Second, we analyze the accuracy of our system and its ability to use resources productively while complying with error bounds. When error bounds are violated, we quantify the number of violations and respective deviations, and, with that, we obtain confidence intervals for error compliance. Finally, we assess the proportion of executions and resources saved for different error bounds. All tests were conducted using 6 machines with an Intel Core i7-2600K CPU at 3.40GHz, 11926MB of RAM memory, and HDD 7200RPM SATA 6Gb/s 32MB cache, connected by 1 Gbps LAN.

Instead of presenting the evaluation for all workloads we experimented, we decided to conduct an in-depth analysis by selecting 2 interesting applications. These applications represent 2 different and realistic scenarios for continuous and incremental processing: i) LRB, a variable tolling system for an urban expressway structure based on the Linear Road Benchmark~\cite{Arasu:2004:LRS:1316689.1316732}; and ii) AQHI, a system based on a network of sensors to classify the quality of the air in a geographic location, inspired by the Air Quality Health Index (AQHI)\footurl{http://www.ec.gc.ca/cas-aqhi/}
used in Canada. 

\begin{figure}
	\centering
	\includegraphics[width=\columnwidth]{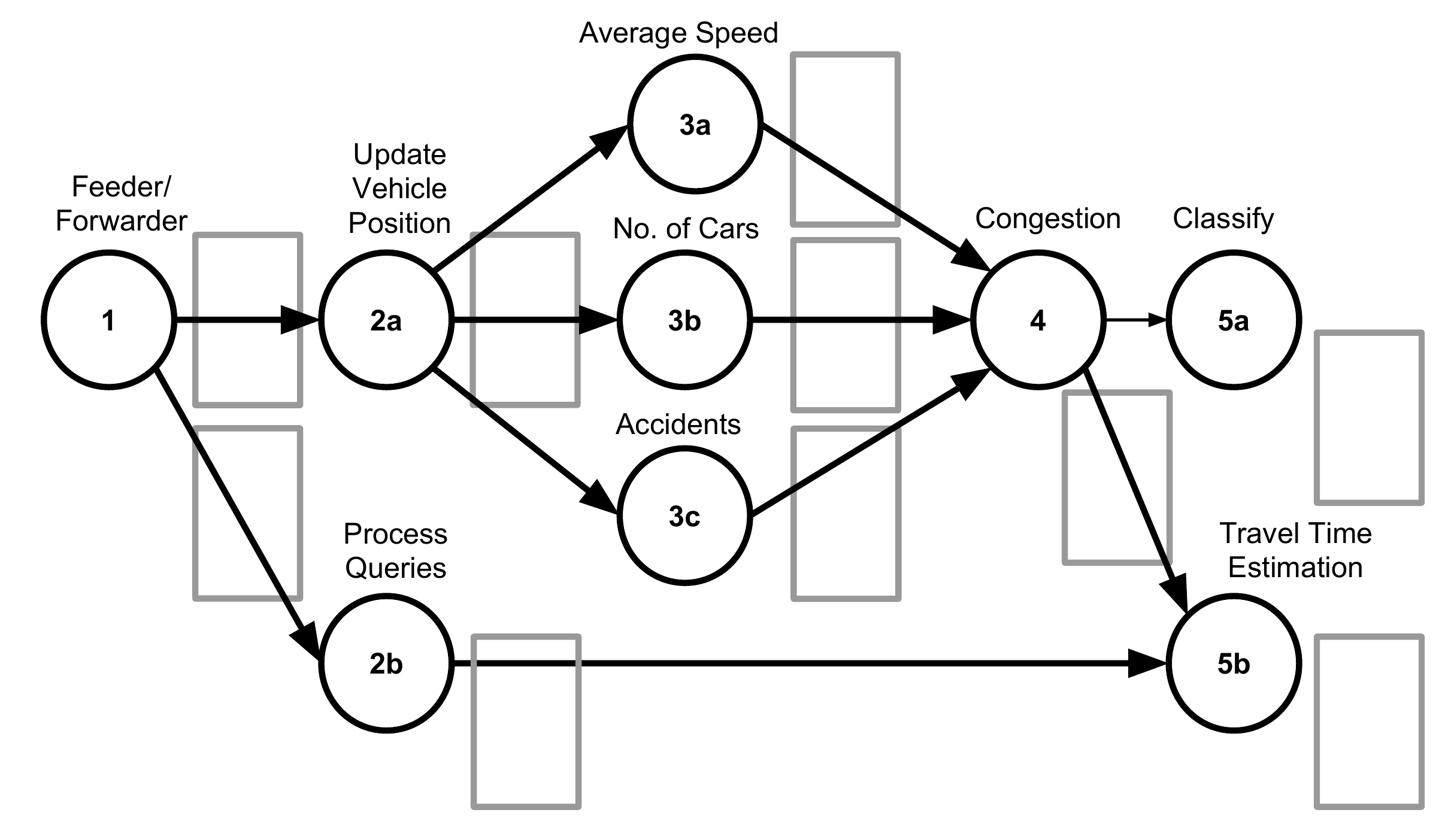}
	\caption{Workflow of the Linear Road. Rectangles in grey represent columnar-like data containers.}
	\label{fig:workflow-lrb}
\end{figure}

\myparagraph{LRB} In the first scenario, we have a variable tolling system for a fictional expressway system where different toll rates are charged based on the time of day or level of congestion of a roadway. The data inputted to the workflow is generated by the MIT-SIMLab (a simulation-based laboratory) \cite{Arasu:2004:LRS:1316689.1316732} and consists of vehicle position reports and historical query requests. Position reports are emitted every 30 seconds by each vehicle, through a transponder, and they identify the vehicle's exact location in the expressway system. Through these reports, we generate statistics comprising average vehicle speed, number of vehicles and existence of accidents, for every segment of every expressway for every minute. Then, these statistics are used to determine toll rates for the segments where the vehicles are in.

Historical query requests, by turn, are issued by vehicles with some small probability every time they emit a position report. Upon such requests, the workflow calculates and reports an account balance, a total of all assessed tolls on a given expressway on a given day, or an estimated travel time and cost for a journey on an expressway.

This tolling system attempts to control the traffic flow by discouraging drivers from using already congested roads, through increased tolls; and, conversely, encouraging the use of less congested roads through decreased tolls. The workflow we designed to process this tolling system is depicted in Figure~\ref{fig:workflow-lrb}. The corresponding processing steps are described as follows.

\textbf{Step 1} receives, separates, and stores position reports and queries from vehicle transponders into different data containers to be processed by step 2a and 2b respectively. \textbf{Step 2a} updates vehicle positions in the urban expressway, which includes updating every segment of every expressway with new vehicle data. This step is only staged to execution when there is a sufficient number of position reports (complying with the QoD of step 2a). \textbf{Steps 3a}, \textbf{3b}, and \textbf{3c} assess the average speed (for all cars in a segment in the last 5 minutes), number of cars, and the existence of accidents on every segment of every expressway respectively. Each of these steps is only triggered when significant differences (according to predefined error bounds) in vehicle positions are perceived against a previous state. Also, the input impact is maintained separately for each of these three steps. \textbf{Step 4} computes the level of congestion for every segment of every expressway based on the average speed and number of vehicles, as well as the presence of accidents nearby. This represents the calculation of the toll in the original benchmark. \textbf{Step 5a} identifies and classifies areas in the expressway system where the traffic congestion is low, medium, or high.
\textbf{Step 2b} processes and prioritizes queries; and \textbf{step 5b} estimates travel time and cost for a journey on an expressway. These 2 last steps are executed synchronously since they generate replies to real time queries.

\begin{figure}
	\centering
	\includegraphics[width=\columnwidth]{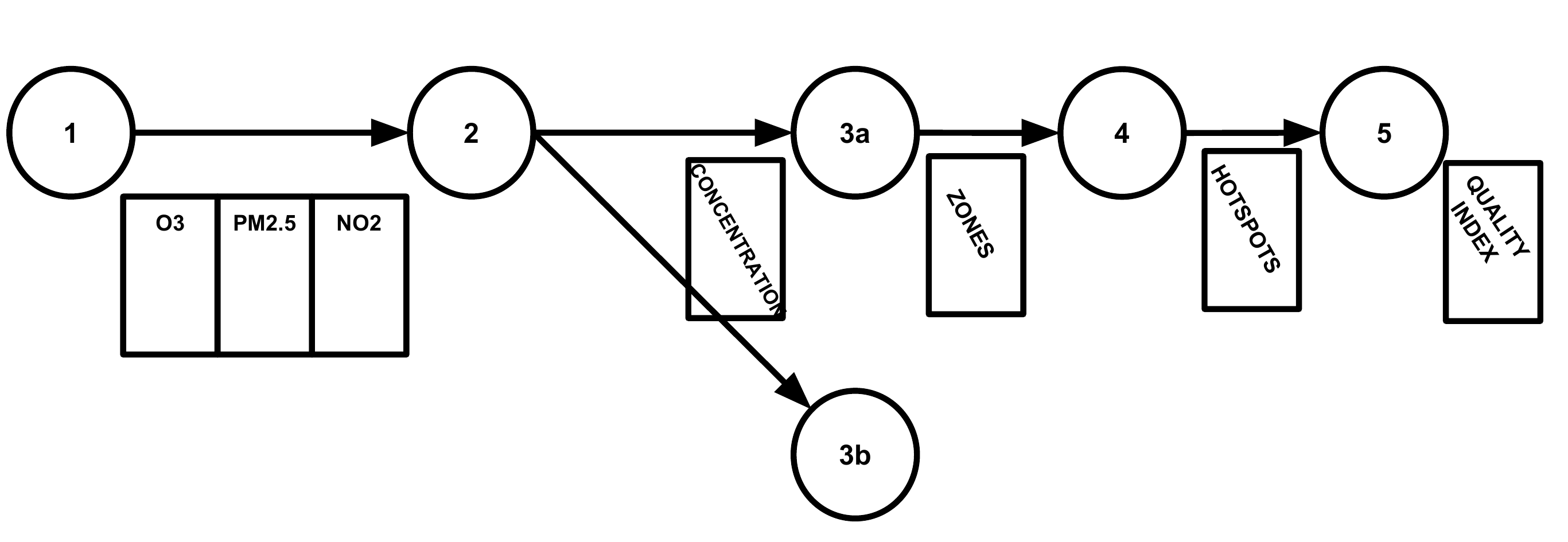}
	\caption{Workflow of the Air Quality Health Index}
	\label{fig:workflow-aqhi}
\end{figure}

\myparagraph{AQHI} Figure~\ref{fig:workflow-aqhi} depicts the workflow that calculates the air quality index (AQHI) on a given geographic region. This index represents a classification of the potential health risk that comes from air pollution. The workflow input is injected from detectors with three sensors to gauge the amount of Ozone ($O_3$), Particulate Matter ($PM_{2.5}$) and Nitrogen Dioxide ($NO_2$) in the atmosphere. In practice, each sensor corresponds to a different generating function, following a distribution with smooth variations across space (i.e., realistic in the variations and trends, while full exactness for a given day record is not relevant for our purposes). These sample variations generated provide the necessary input data to the workflow in each (re-)execution, a wave, corresponding to an hour of the day, for a total of 168 waves for a full week simulated. The generating functions return a value from 0 to 100, where 0 and 100 are, respectively, the minimum and maximum known values of $O_3$, $PM_{2.5}$ and $NO_2$. The workflow output corresponds to the generation of an index, a number, that is mapped into a class of health risk: low (1-3), moderate (4-6), high (7-10), and very high (above 10).


\textbf{Step 1} simulates asynchronous and deferred arrival of sensory data. It continuously receives data from the atmospheric sensors and feeds the workflow by updating the first data container (composed of 3 columns). \textbf{Step 2} calculates a single value, through a multiplicative model, representing the combined concentration of the 3 sensors for each detector. Every single calculated value is written on column concentration of the data store. \textbf{Step 3a} divides the considered region in smaller areas and computes the aggregated concentration of pollution from detectors in each area. \textbf{Step 3b} processes the concentration of the area between detectors, thereby averaging the concentration perceived by surrounding detectors. It also plots a chart containing a representation of the pollution concentrations throughout the whole probed area for displaying purposes. \textbf{Step 4} assesses which of the previous stored zones have a concentration above a specified reference, which represents a point from which a zone is considered an hotspot (i.e., zone exhibiting an high level of pollution). \textbf{Step 5} reasons about the hotspots previously detected and, through a simple additive model that combines the number of hotspots with the average concentration of pollution on hotspots, it calculates an index that classifies the overall level of pollution in the given geographic region.

\myparagraph{Correlation between Input Impact and Error}
\begin{figure*}
    \centering
    \subfigure[LRB(3a) Average Speed] {
    \includegraphics[width=0.23\textwidth]{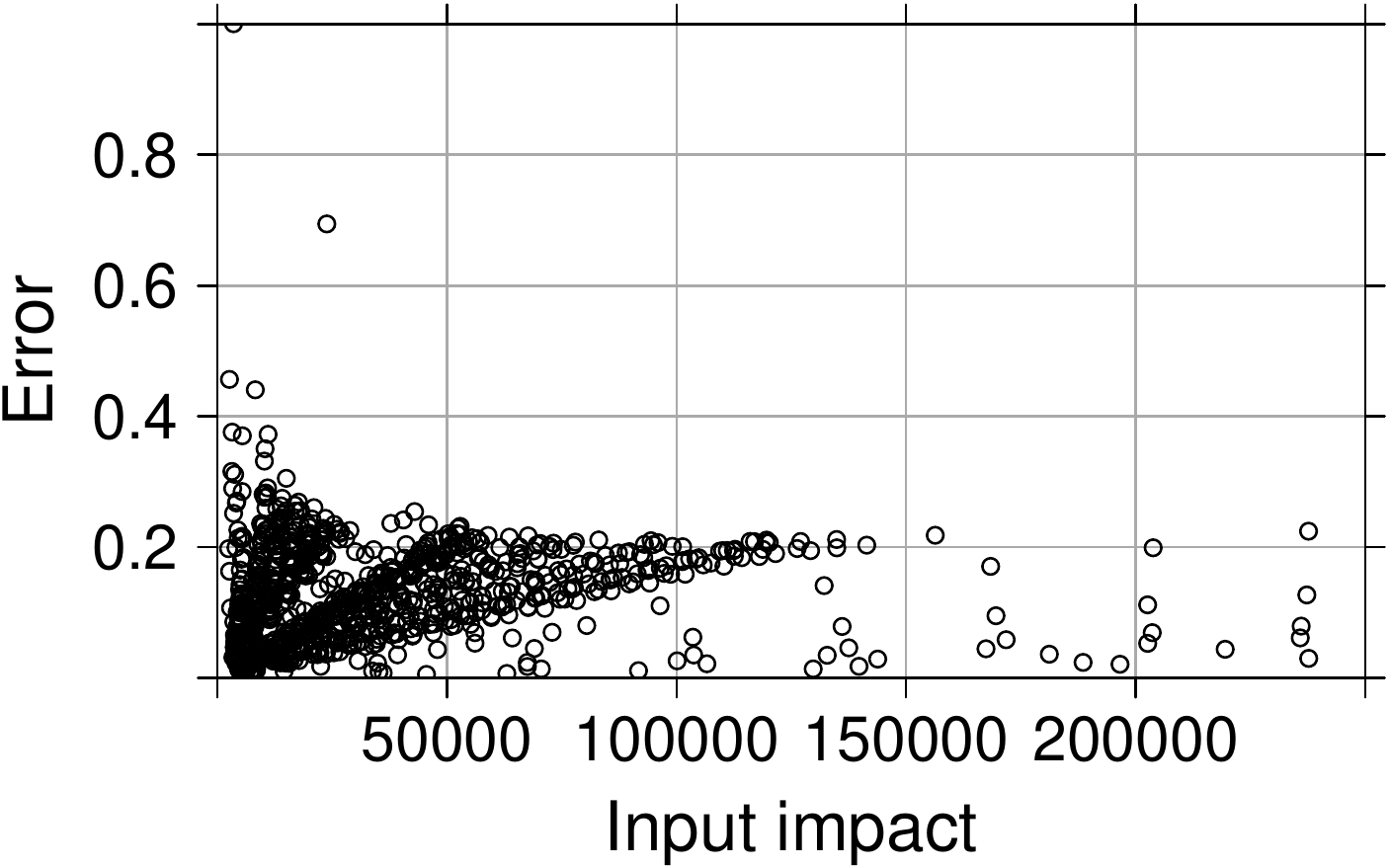}
    \label{fig:correlation-avg-speed}
    }
    \subfigure[LRB(3b) Number of Cars] {
    \includegraphics[width=0.23\textwidth]{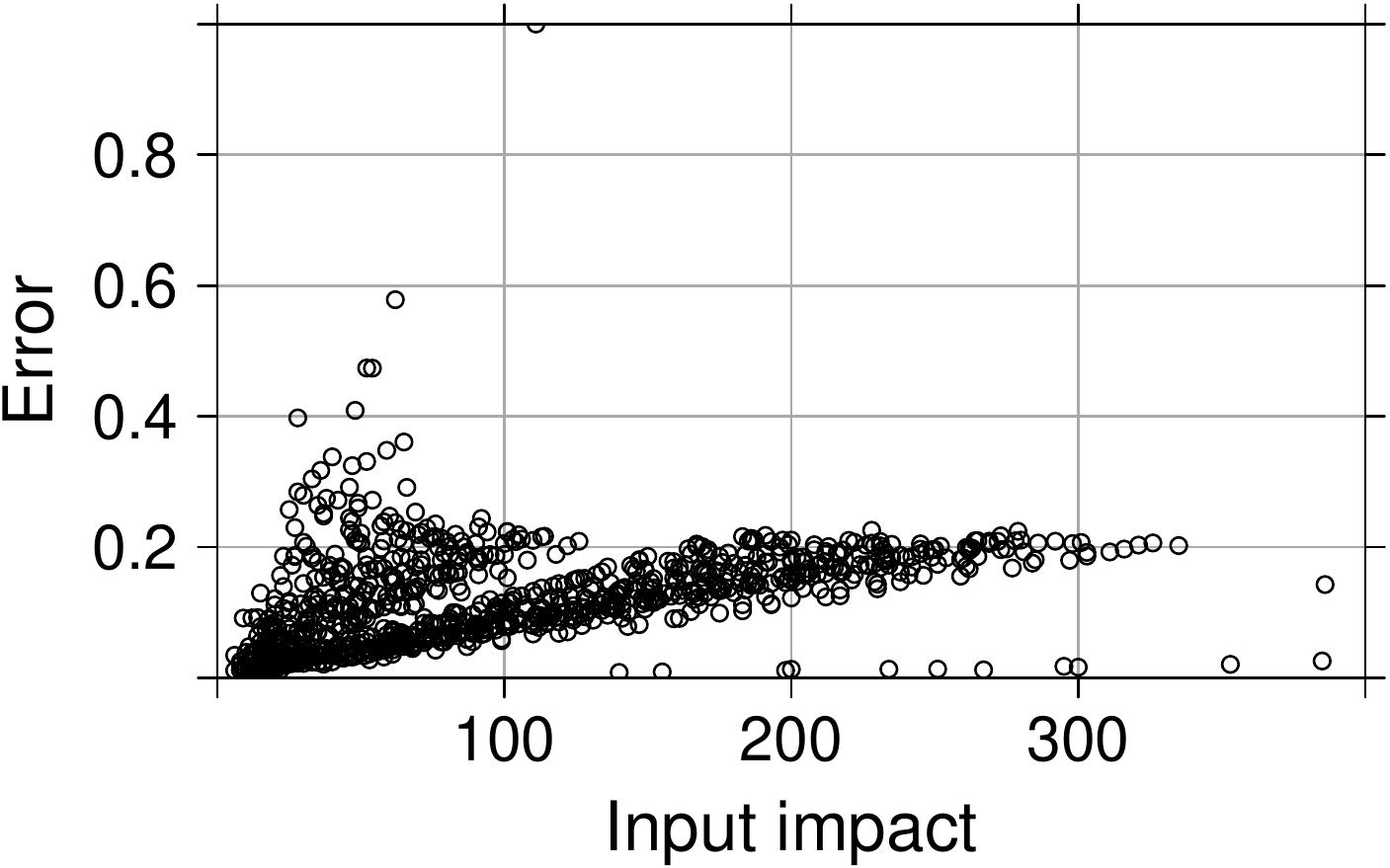}
    \label{fig:correlation-no-of-cars}
    }
    \subfigure[LRB(4) Congestion] {
    \includegraphics[width=0.23\textwidth]{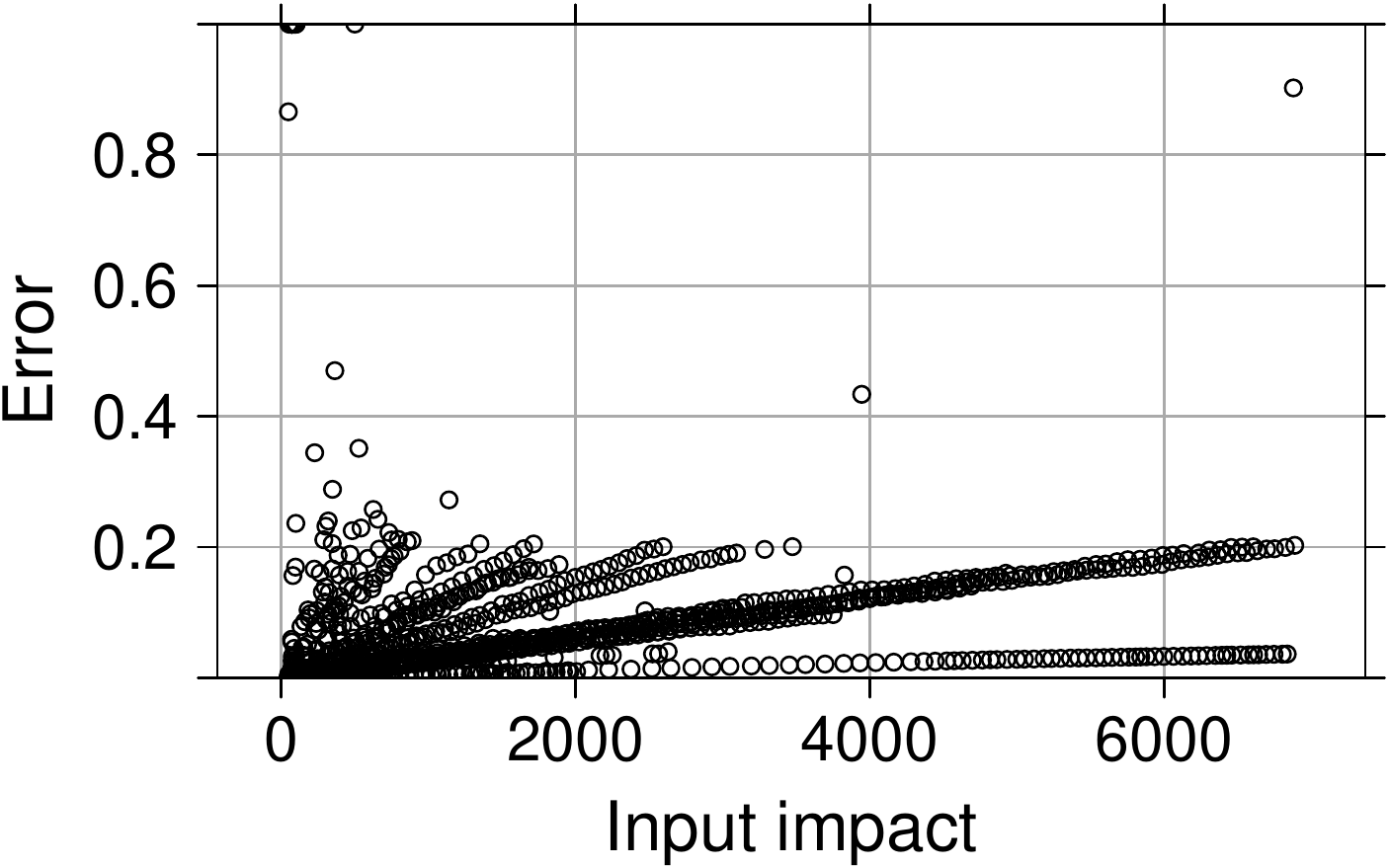}
    \label{fig:correlation-congestion}
    }
    \subfigure[LRB(5a) Classify] {
    \includegraphics[width=0.23\textwidth]{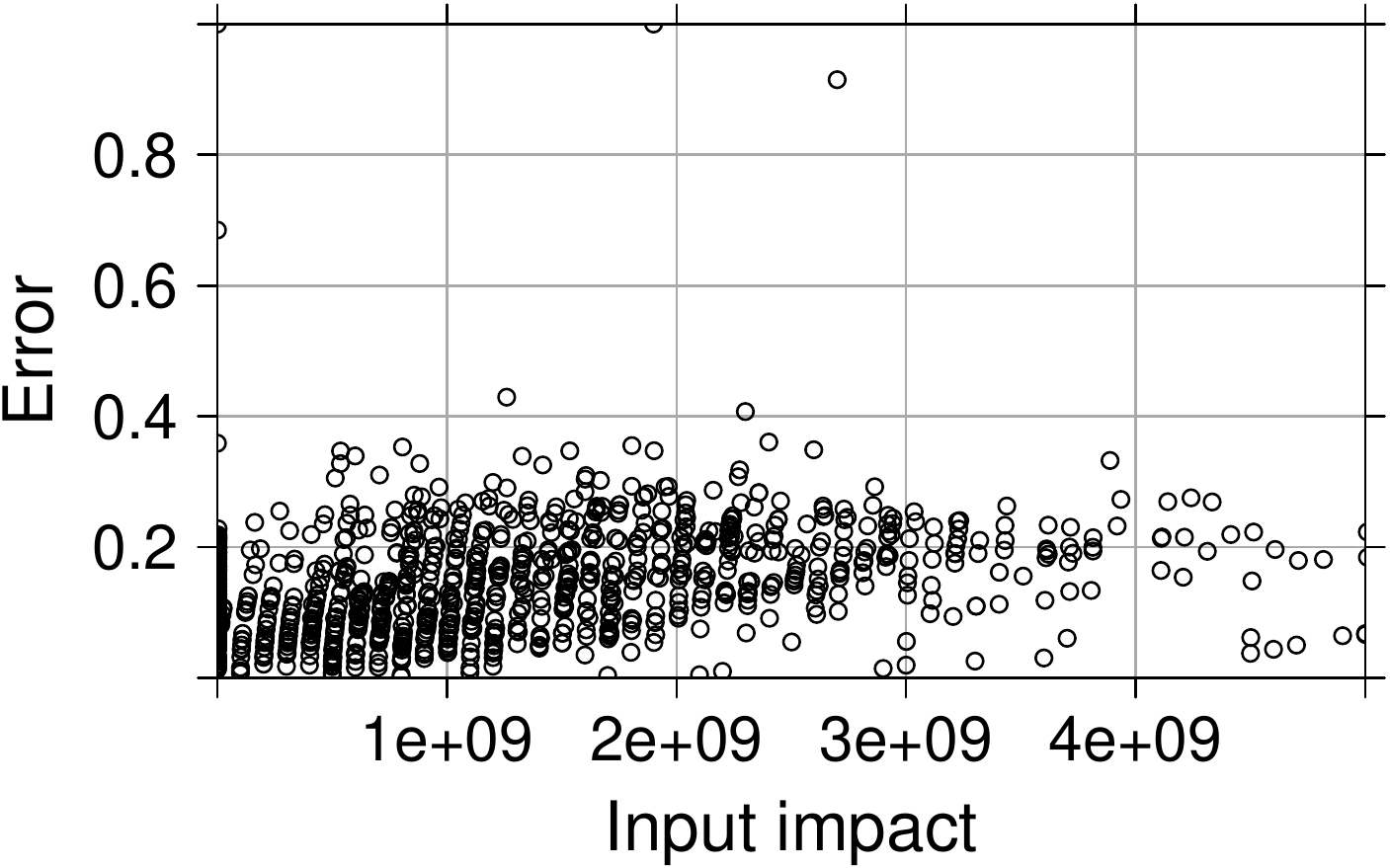}
    \label{fig:correlation-classify}
    }
    \\
    \subfigure[AQHI(3a) Zones] {
    \includegraphics[width=0.23\textwidth]{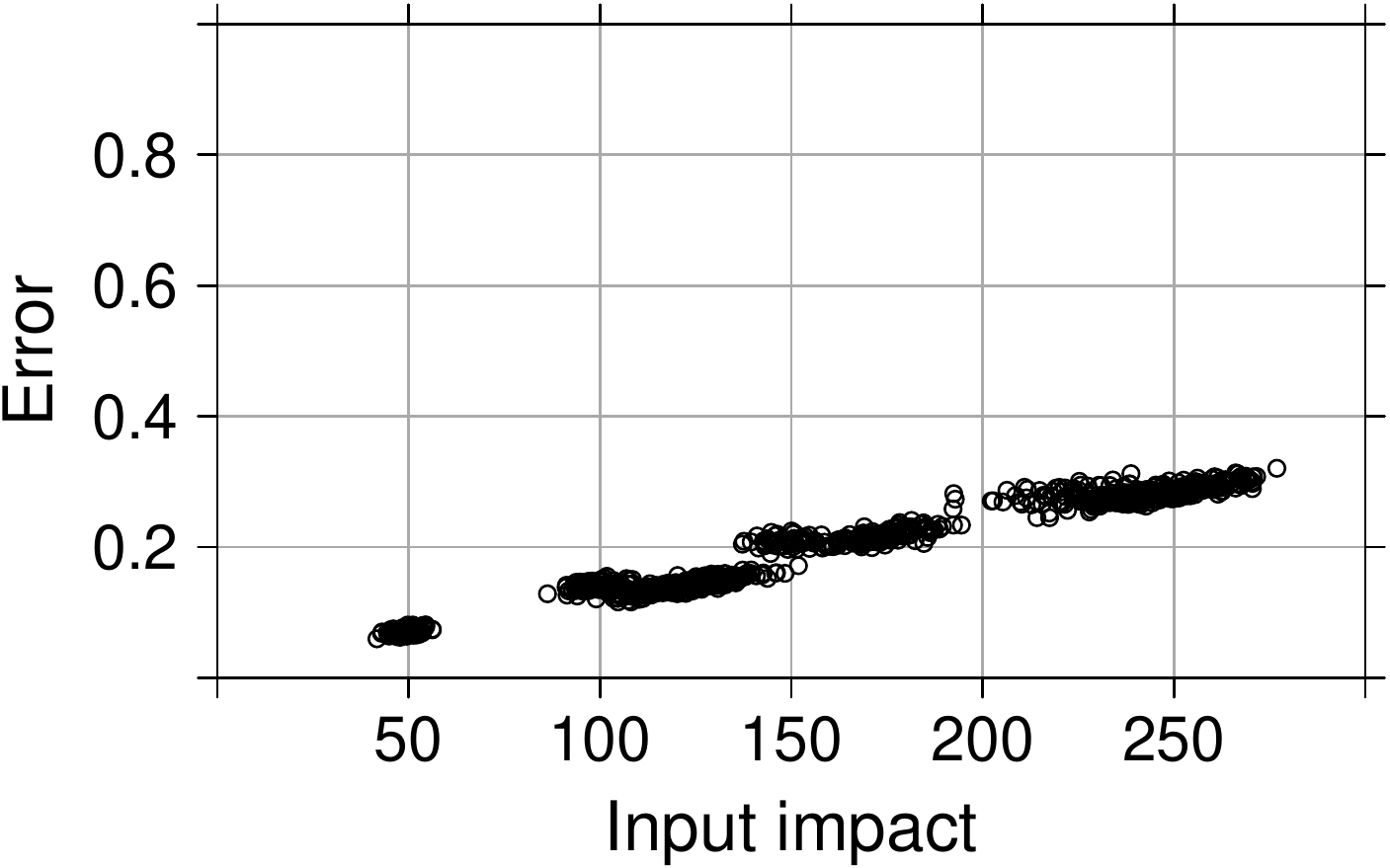}
    \label{fig:correlation-zones}
    }
    \subfigure[AQHI(4) Hotspots] {
    \includegraphics[width=0.23\textwidth]{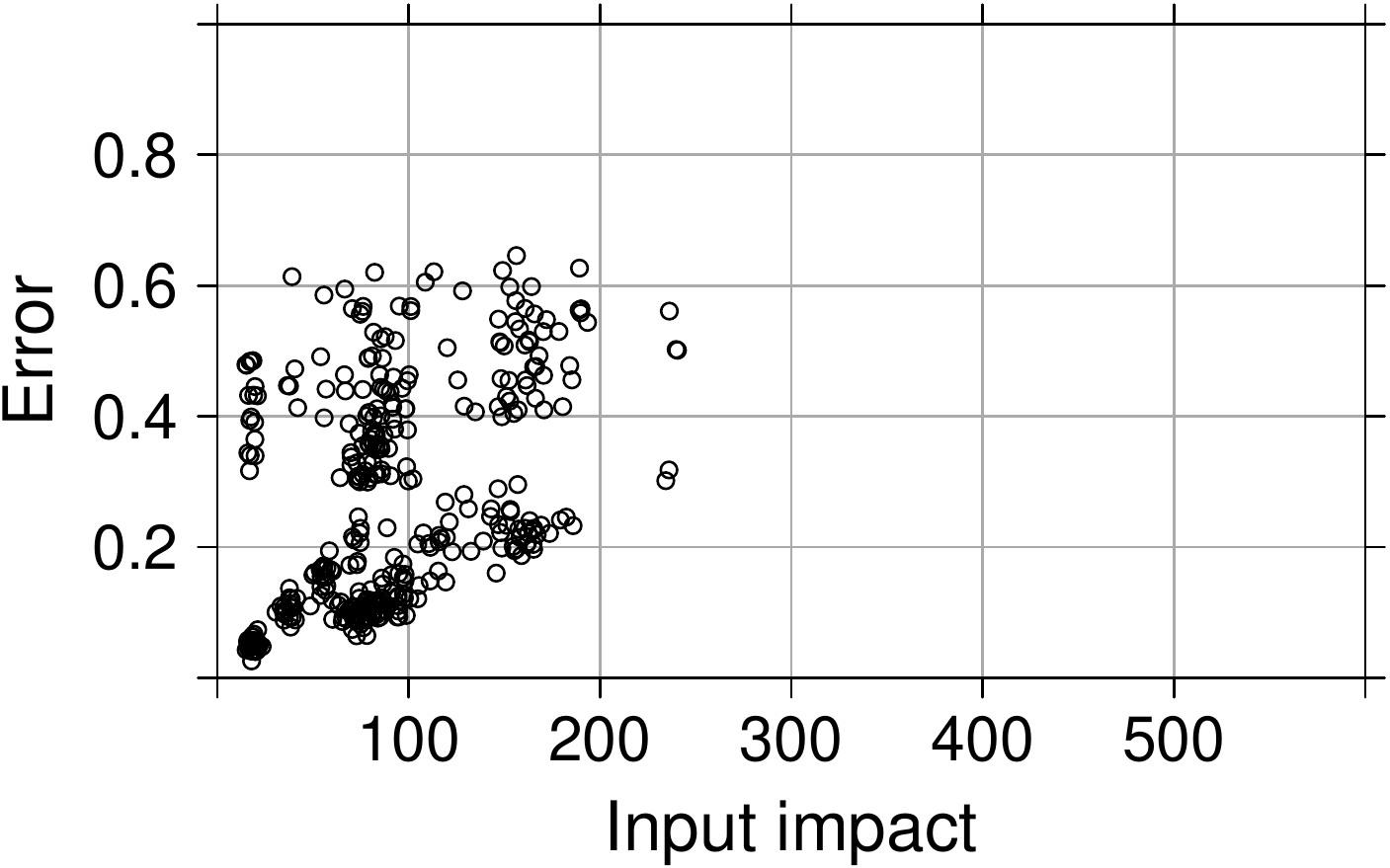}
    \label{fig:correlation-hotspots}
    }
    \subfigure[AQHI(5) AQHI] {
    \includegraphics[width=0.23\textwidth]{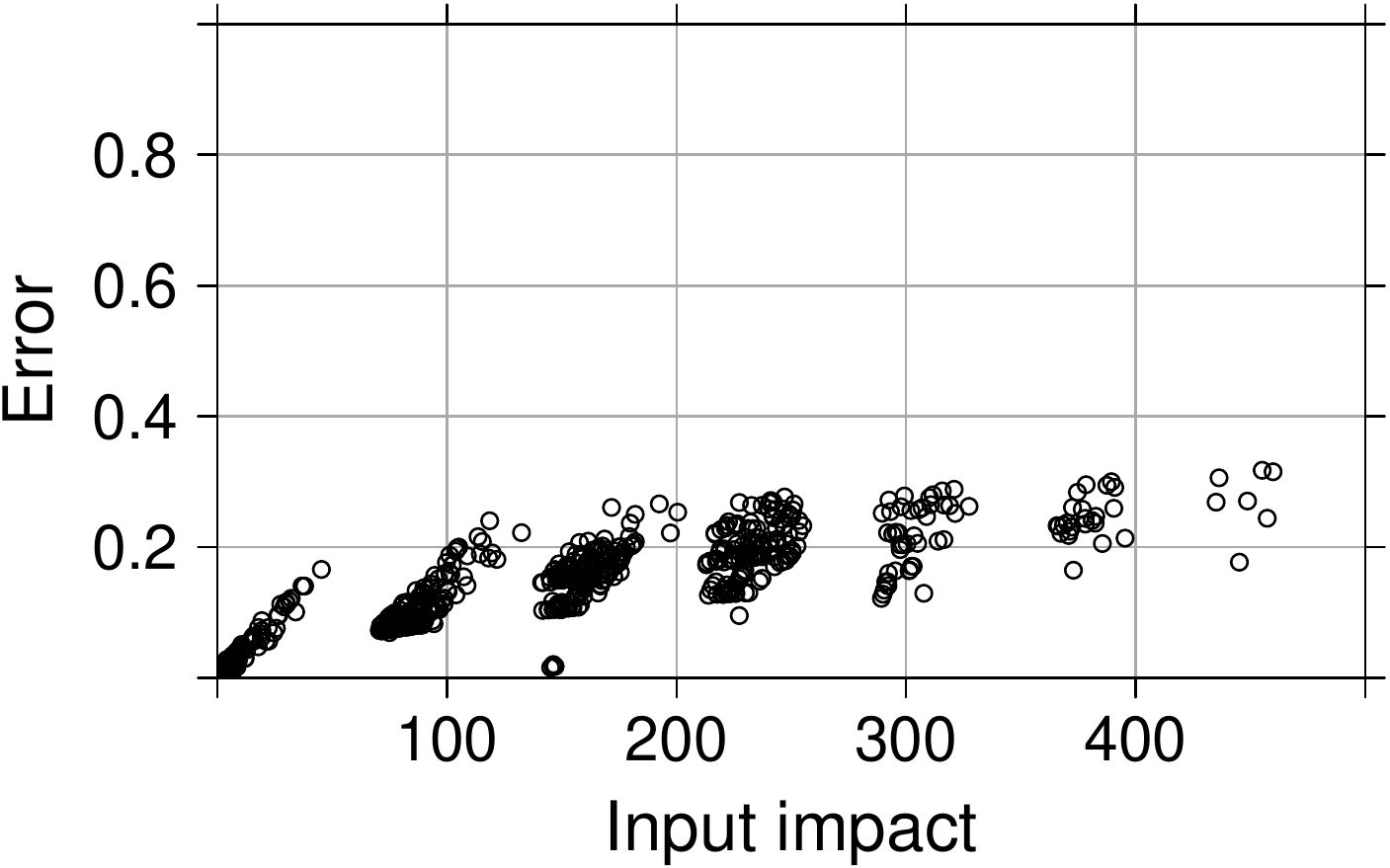}
    \label{fig:correlation-aqhi}
    }

    \caption{Correlation between input impact and error for the main processing steps of LRB and AQHI}
    \label{fig:correlation-in-error}
\end{figure*}
Figure~\ref{fig:correlation-in-error} shows the correlation between input impact and error for the main processing steps of LRB (figures~\ref{fig:correlation-avg-speed}-\ref{fig:correlation-classify}) and AQHI (\ref{fig:correlation-zones}-\ref{fig:correlation-aqhi}), using a maximum tolerated error of 20\%. These are the steps that tolerate error and that exhibit the most interesting patterns. We can observe that the correlations vary across steps and that most of them are neither linear nor trivial to be simply deduced. If they were obvious, other simpler techniques like linear regression would suffice. Hence, we justify the use of Machine Learning to learn these complex patterns, that vary according to the computations being performed, and ensure the error is bounded. From all the figures, \ref{fig:correlation-classify} and~\ref{fig:correlation-hotspots} exhibit higher variance, which carries more complexity to the learning process. Nevertheless, as long as these patterns are learned during a training phase, RF does a very good job in recognizing them (in the future) as we can see in the following experiments.

\myparagraph{Prediction Accuracy}
\begin{figure*}[ht!]
    \centering

    \subfigure[LRB Accuracy] {
    	\includegraphics[width=0.25\textwidth]{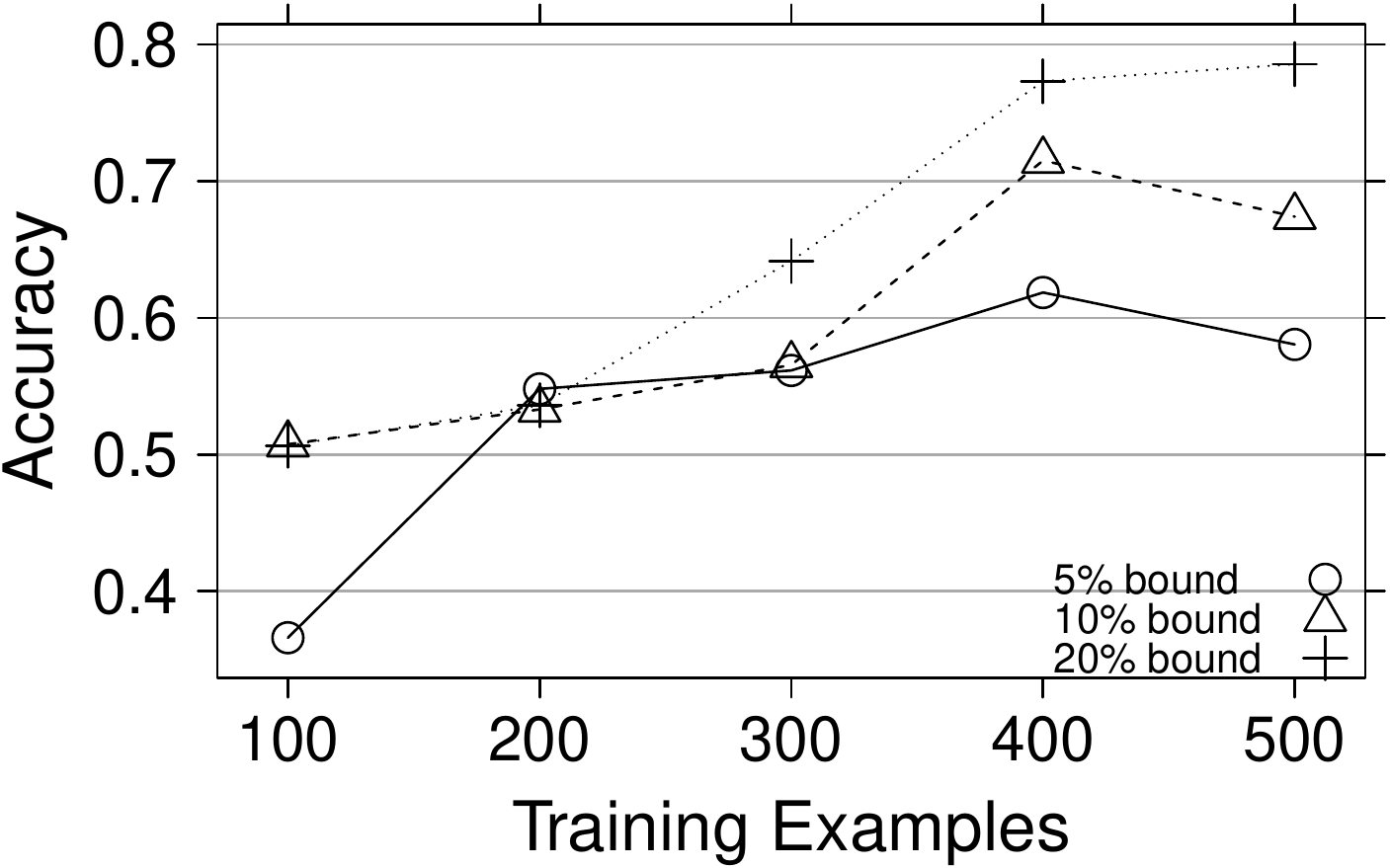}
    	\label{fig:accuracy-2d-a}
    }
	\subfigure[LRB Precision] {
		\includegraphics[width=0.25\textwidth]{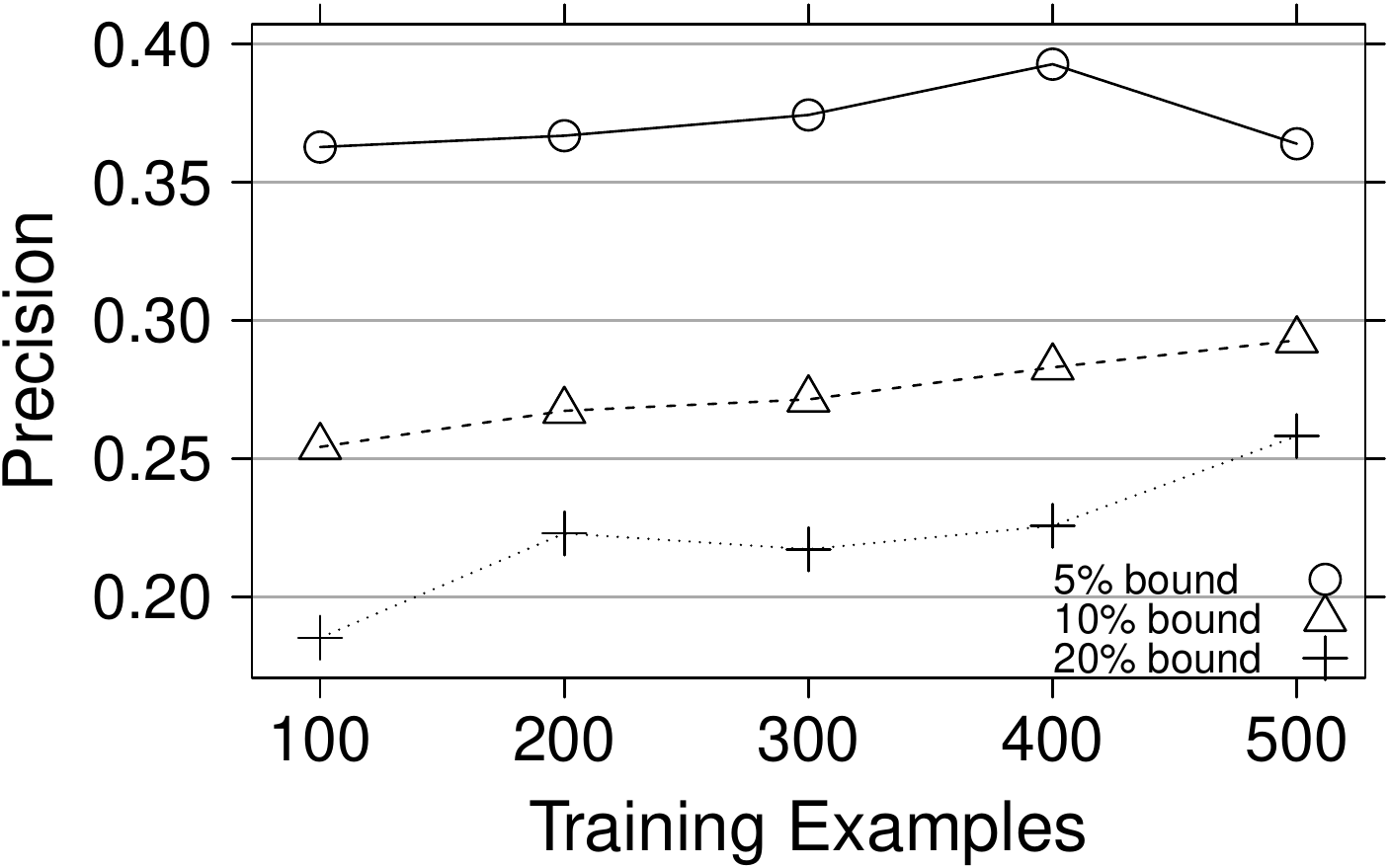}
    	\label{fig:accuracy-2d-b}		
	}
    \subfigure[LRB Recall] {
    	\includegraphics[width=0.25\textwidth]{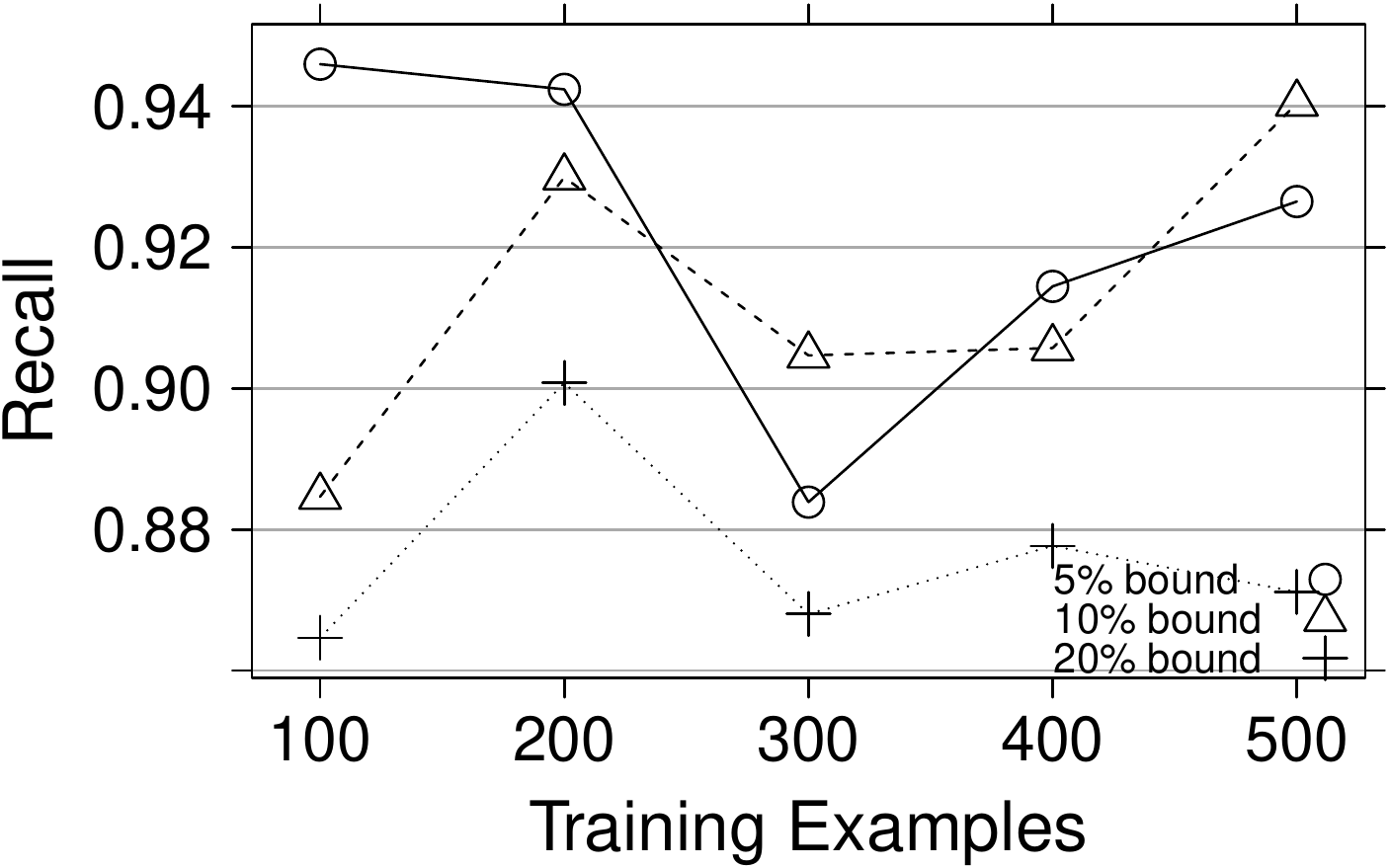}
    	\label{fig:accuracy-2d-c}
    }\\

    \subfigure[AQHI Accuracy] {
    	\includegraphics[width=0.25\textwidth]{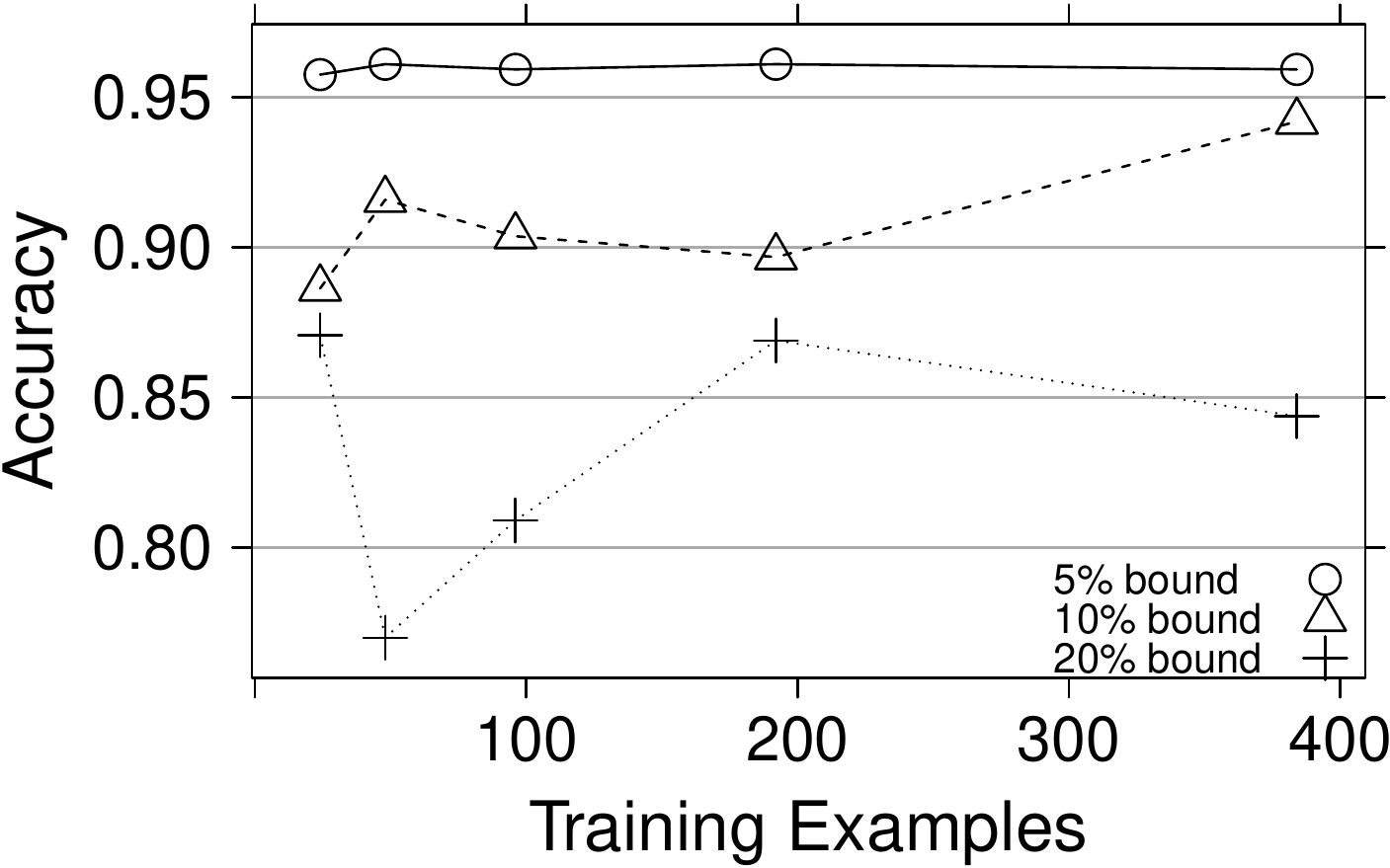}
    	\label{fig:accuracy-2d-d}
    }
	\subfigure[AQHI Precision] {
		\includegraphics[width=0.25\textwidth]{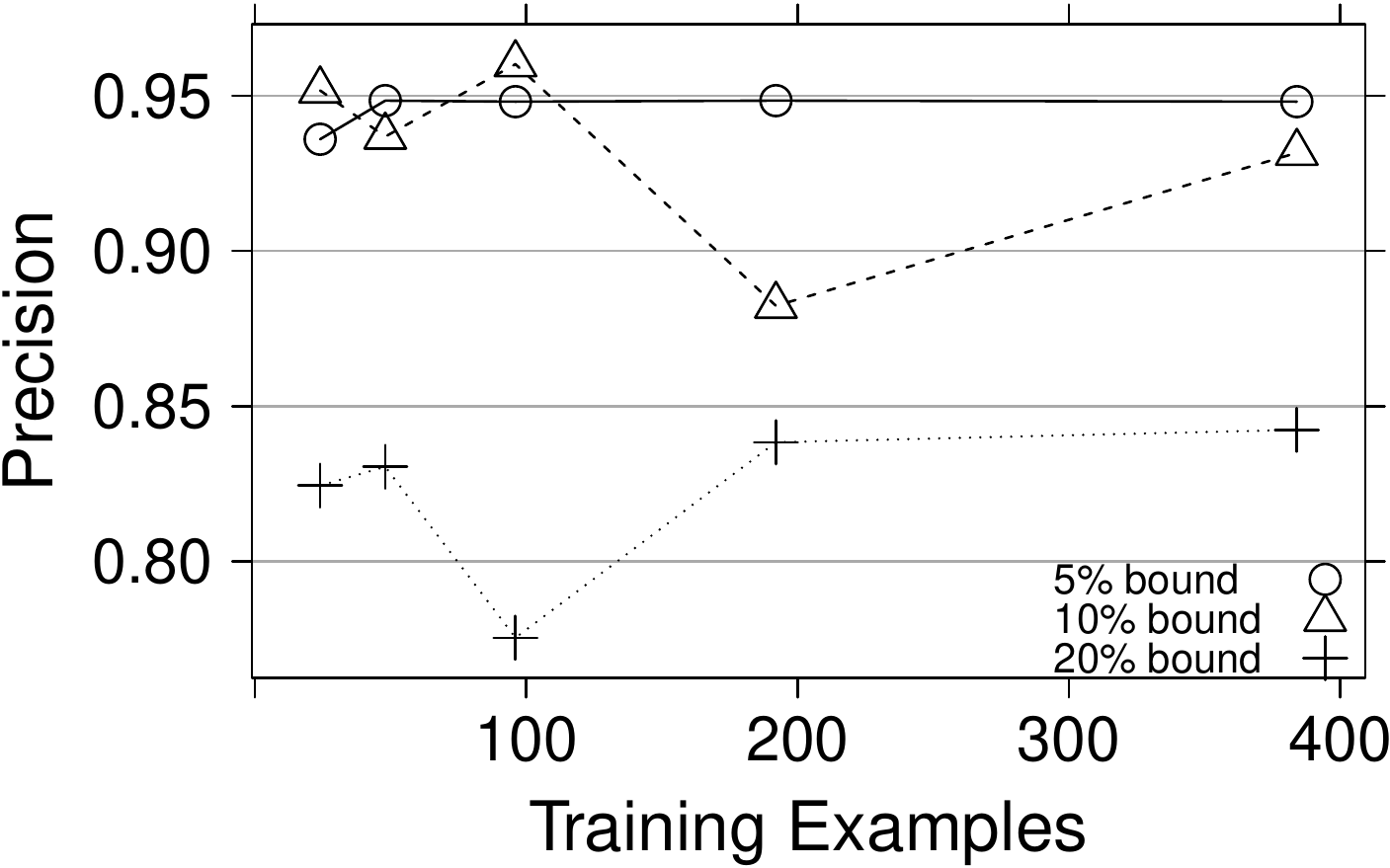}
    	\label{fig:accuracy-2d-e}		
	}
    \subfigure[AQHI Recall] {
    	\includegraphics[width=0.25\textwidth]{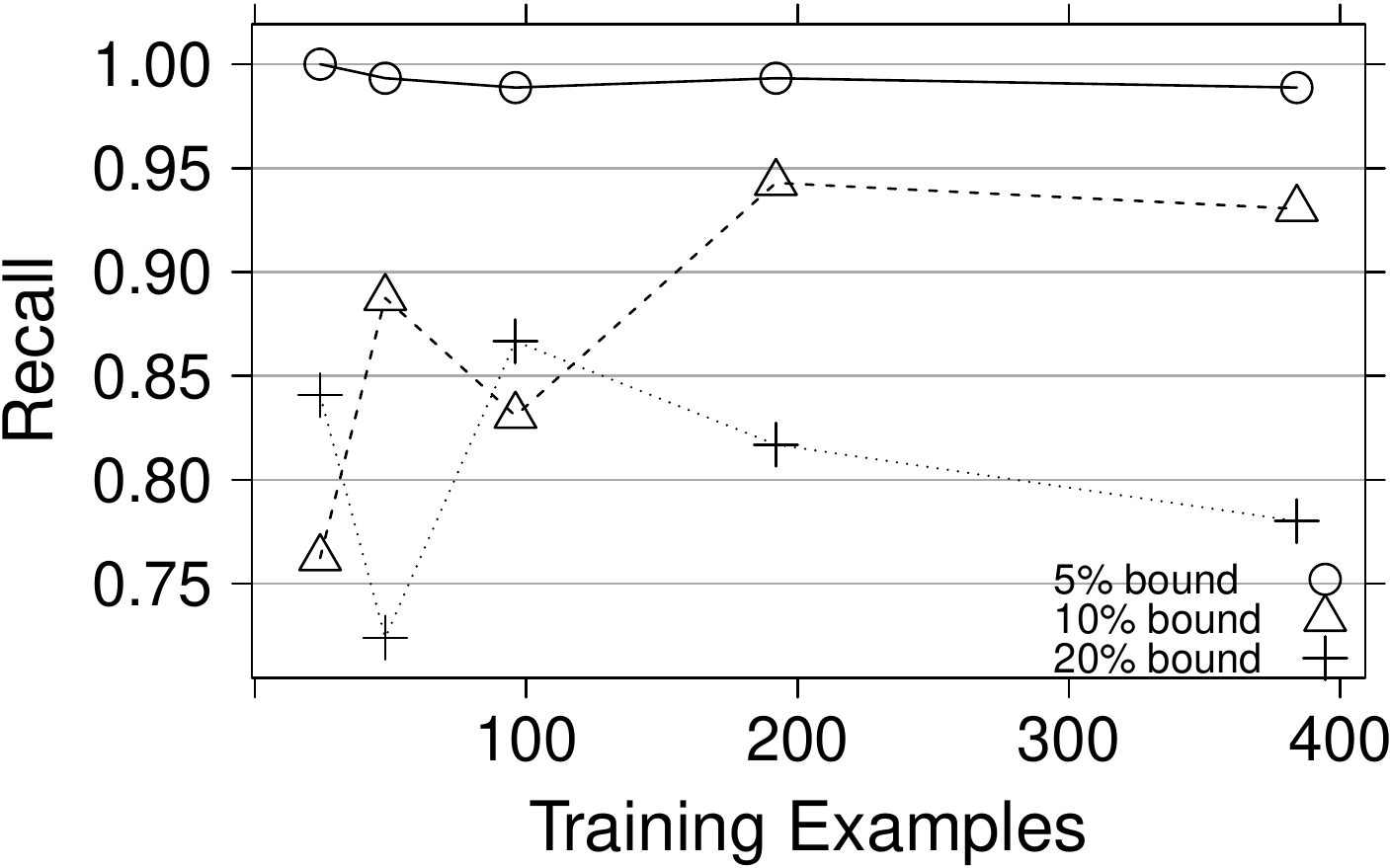}
    	\label{fig:accuracy-2d-f}
    }

    \caption{Accuracy, Precision, and Recall for LRB and AQHI with error bounds of 5, 10, and 20\%}
    \label{fig:accuracy-2d}
\end{figure*}
Figure~\ref{fig:accuracy-2d} shows the accuracy, precision, and recall,
while varying the number of examples in the training-set, for LRB and AQHI using error bounds of 5, 10, and 20\%. The examples contained in the test-sets were taken in subsequent waves as those of training-sets. 500 test examples were used for LRB and 384 for AQHI (respectively corresponding to a cycle of a pattern that repeats across time). Since LRB exhibited more variance, we decided to optimize its classifier for recall, minimizing error deviations above $max_\varepsilon$.

For the LRB, subfigures~\ref{fig:accuracy-2d-a}-\ref{fig:accuracy-2d-c}, we may observe that the accuracy improves as the number of training examples and error bound increase, up to 80\% when $max_\varepsilon=20\%$. This indicates that with 500 examples our learning model was able to predict execution of steps in 60 upto 80\% of the times in an optimal manner. Optimal means that $max_\varepsilon$ was never exceeded and step re-execution was postponed as much as it was possible. However, not having a fully accurate model does not mean that $max_\varepsilon$ is exceeded; e.g., re-execution can happen one wave before the ideal one, preventing $max_\varepsilon$ from being reached, but also leaving space for one execution that could have been saved. We may also notice that the recall is always above 86\% for more than 300 examples in the training-set, meaning that false negatives were reduced and true negatives augmented (i.e., maximizing $max_\varepsilon$ compliance). As a consequence of optimizing for recall, we also get more false positives (less saved executions), which is represented by the precision metric.

As for the AQHI, figures~\ref{fig:accuracy-2d-d}-\ref{fig:accuracy-2d-f}, we may observe that, with a bound of 5\%, all metrics yield values equal or higher than 95\%, which constitutes an excellent result (i.e., almost optimal resource savings and error compliance). The main reason for this is that the error variation, from wave to wave, was most of the time above 5\% for the first 2 steps, which caused their re-execution in almost every wave. For an error bound of 10\%, accuracy was roughly stable across different training-sets, and above 90\% for more than 100 examples in training-set. Recall increased with the number of training examples upto roughly 100\%, showing that $max_\epsilon$ was almost never violated. Conversely, precision slightly decreased with the number of examples, showing that steps were re-executed more than the ideal necessary to stay within error limits. Finally, for $max_\varepsilon = 20\%$, there is an initial accentuated decline for accuracy and recall until roughly 100 training examples, probably corresponding to less than a complete pattern cycle. After which, accuracy goes from roughly 80 to 90\%, and recall from 80 to 100\%. As expected, AQHI is more stable than LRB. There is more bias and less variability in the input data, changing overall more smoothly cross time. Therefore, the classifier requires less training examples to perform accurate predictions on new unseen examples. Intuitively, the higher the bound (i.e., the \emph{slack} we allow for data modification over time), the higher potential for saving resources, but the less ability to avoid large deviations in the outcome of the execution.


\myparagraph{Measured versus Predicted Errors}
\begin{figure*}[t!]
    \centering

    \subfigure[LRB(5\%) Error] {
    	\includegraphics[width=.25\textwidth]{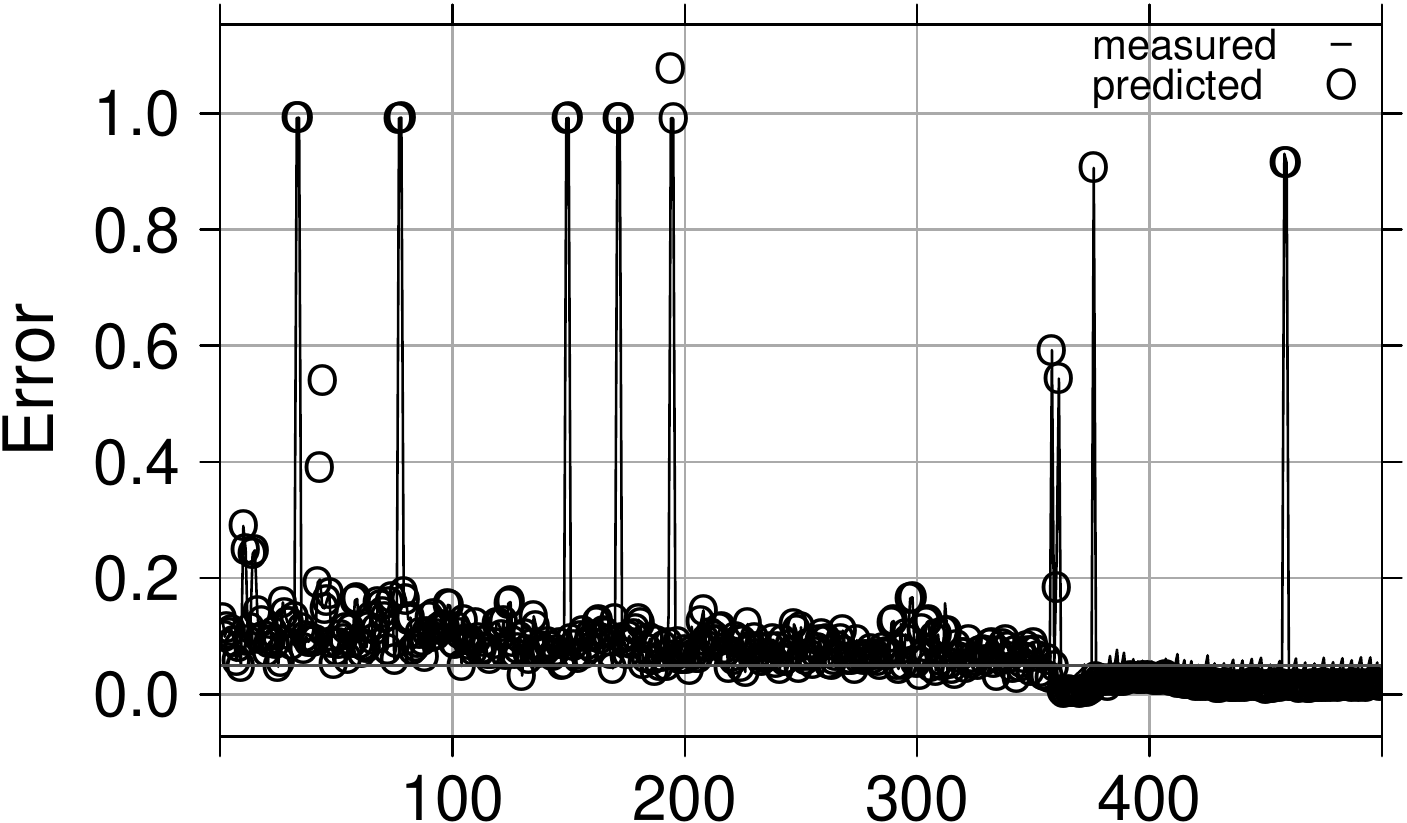}
    	\label{fig:error-a}
    }
    \subfigure[LRB(10\%) Error] {
    	\includegraphics[width=.25\textwidth]{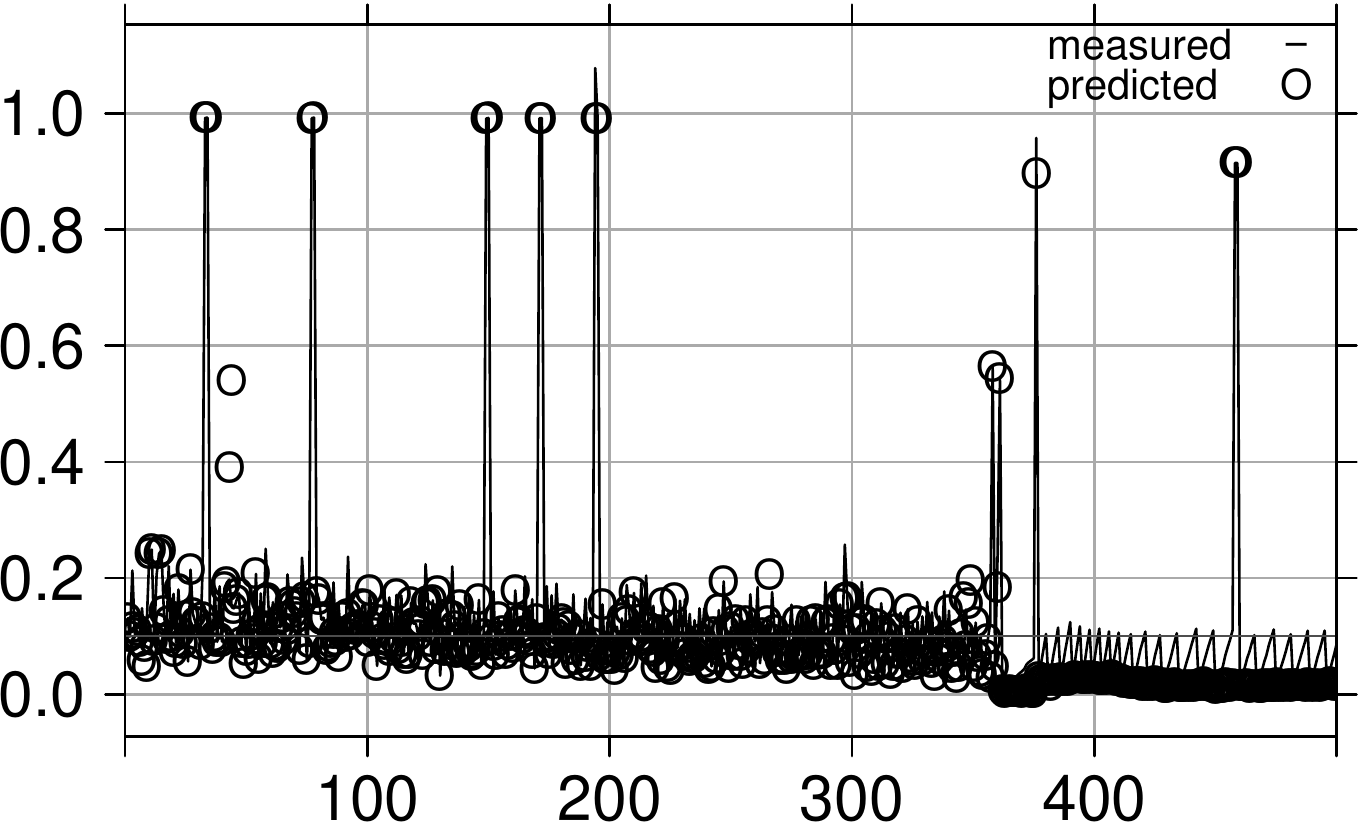}
    	\label{fig:error-b}
    }
    \subfigure[LRB(20\%) Error] {
    	\includegraphics[width=.25\textwidth]{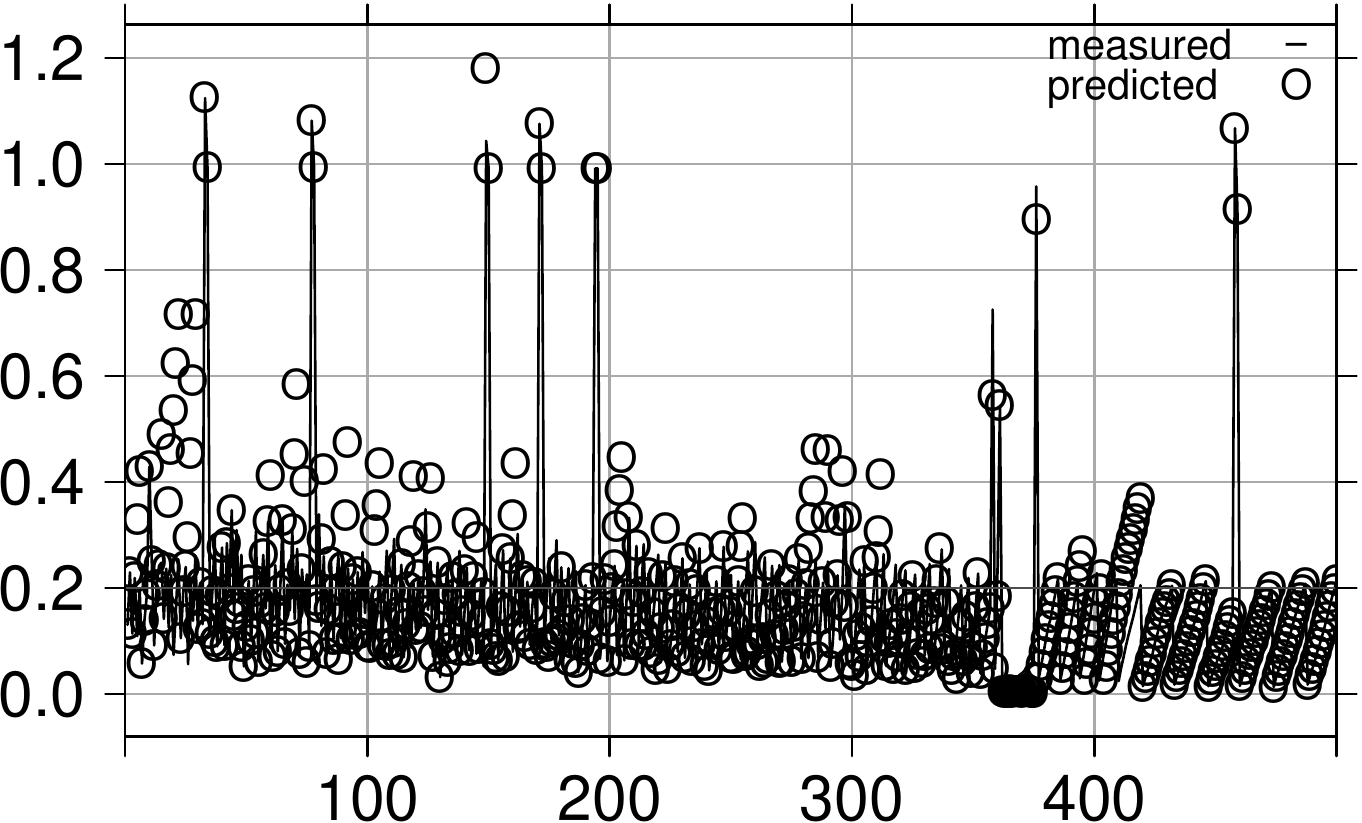}
    	\label{fig:error-c}
    }\\
        
 	\subfigure[LRB(5\%) Prediction Deviation] {
    	\includegraphics[width=0.25\textwidth]{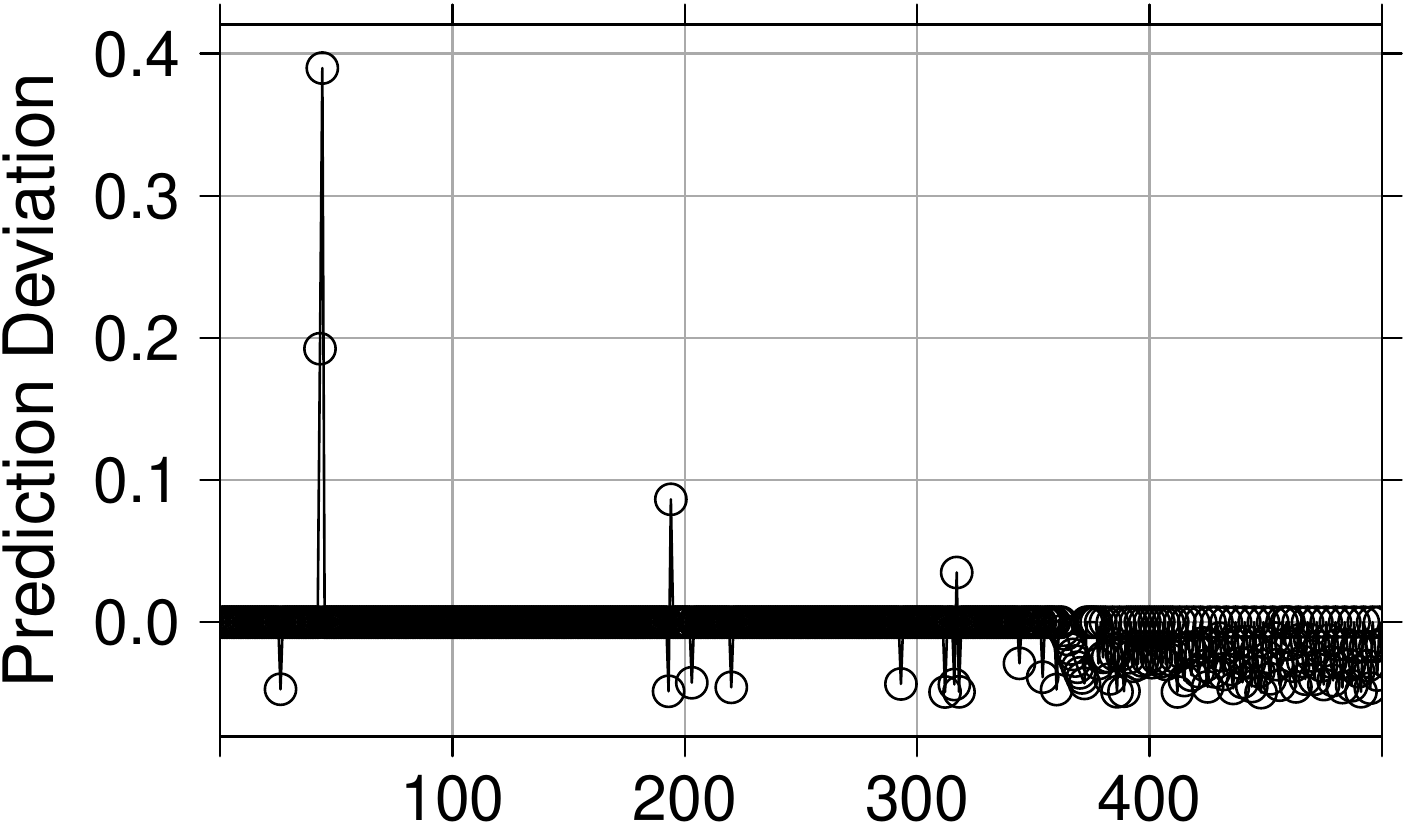}
    	\label{fig:error-d}
    }
 	\subfigure[LRB(10\%) Prediction Deviation] {
    	\includegraphics[width=0.25\textwidth]{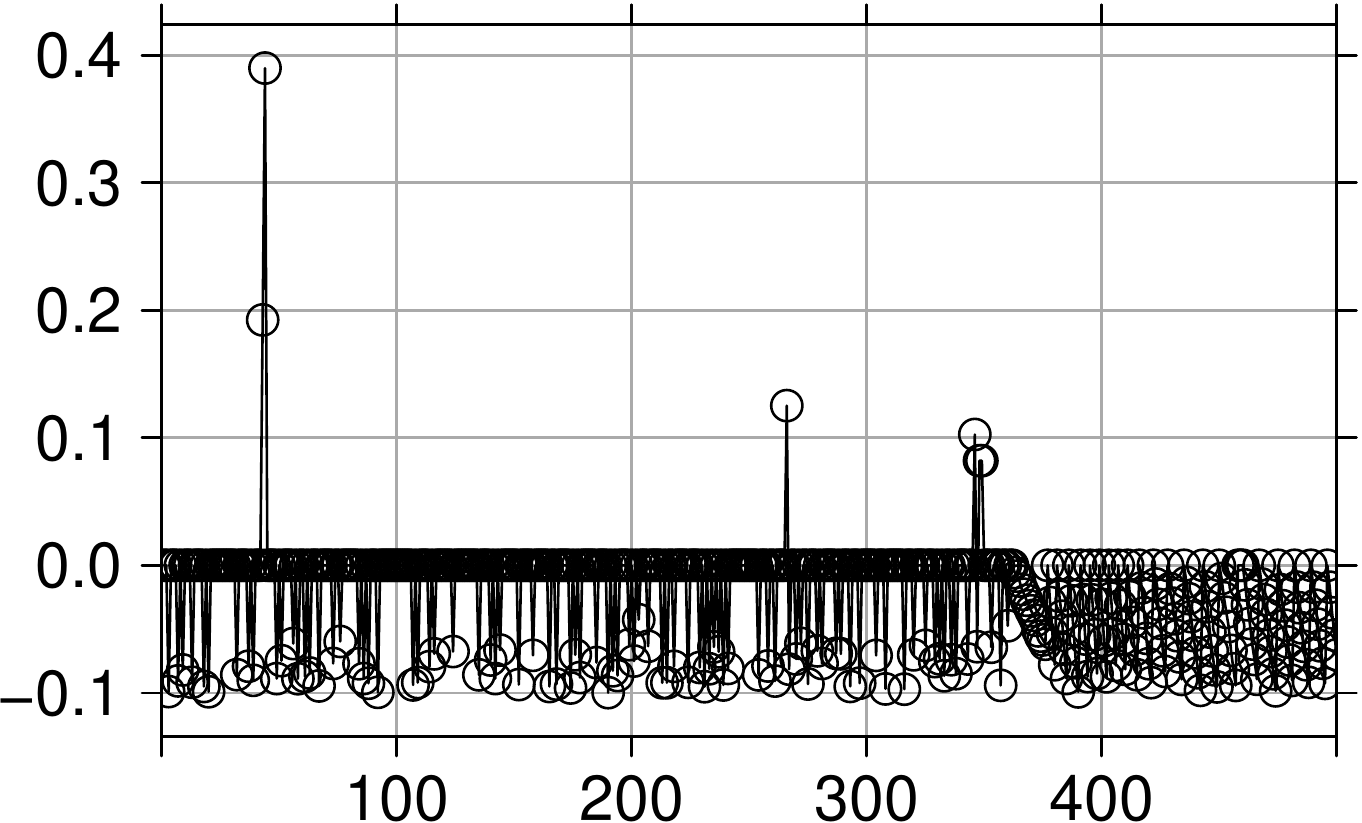}
    	\label{fig:error-e}
    }    
 	\subfigure[LRB(20\%) Prediction Deviation] {
    	\includegraphics[width=0.25\textwidth]{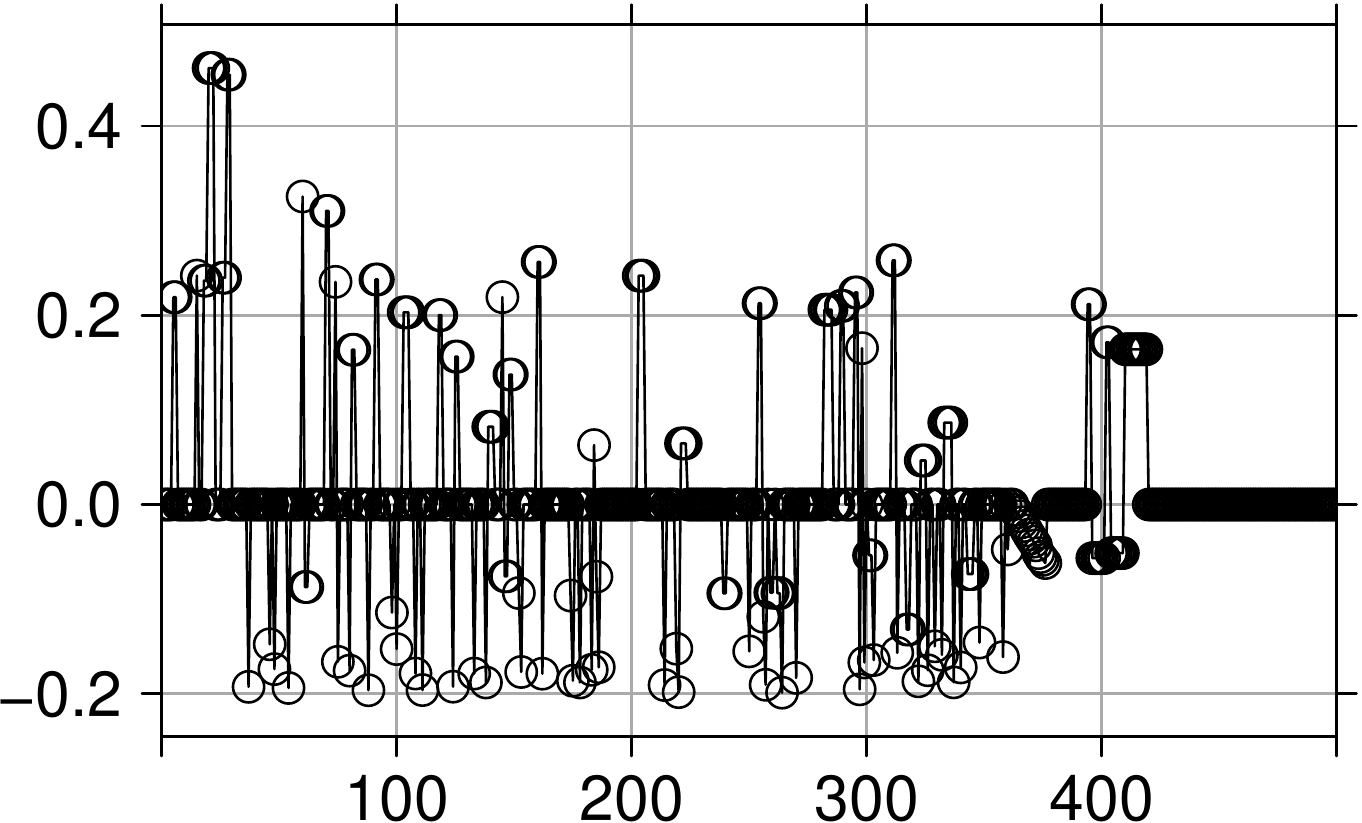}
    	\label{fig:error-f}
    } \\


    \subfigure[AQHI(5\%) Error] {    
    	\includegraphics[width=0.25\textwidth]{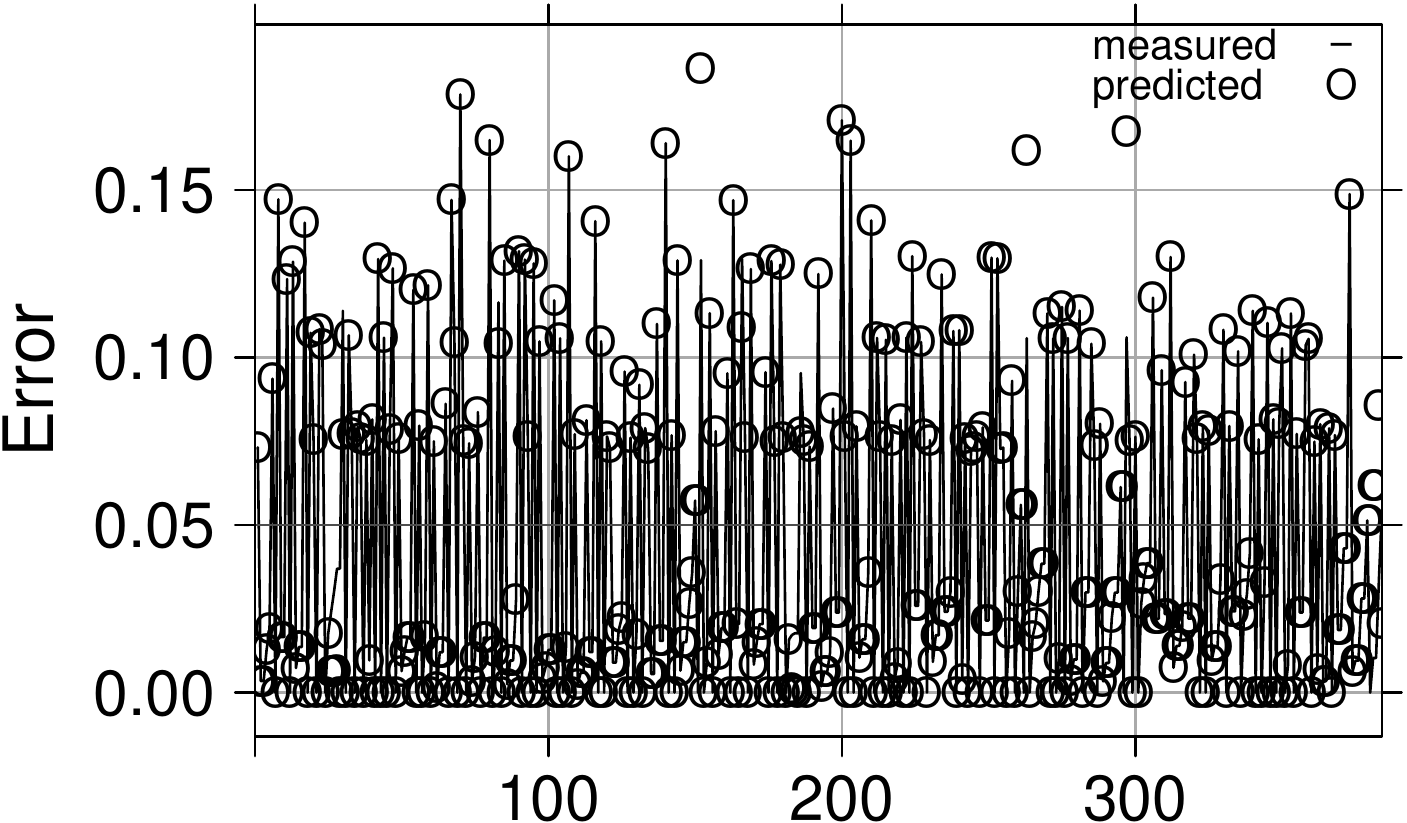}
    	\label{fig:error-g}
    }
    \subfigure[AQHI(10\%) Error] {    
    	\includegraphics[width=0.25\textwidth]{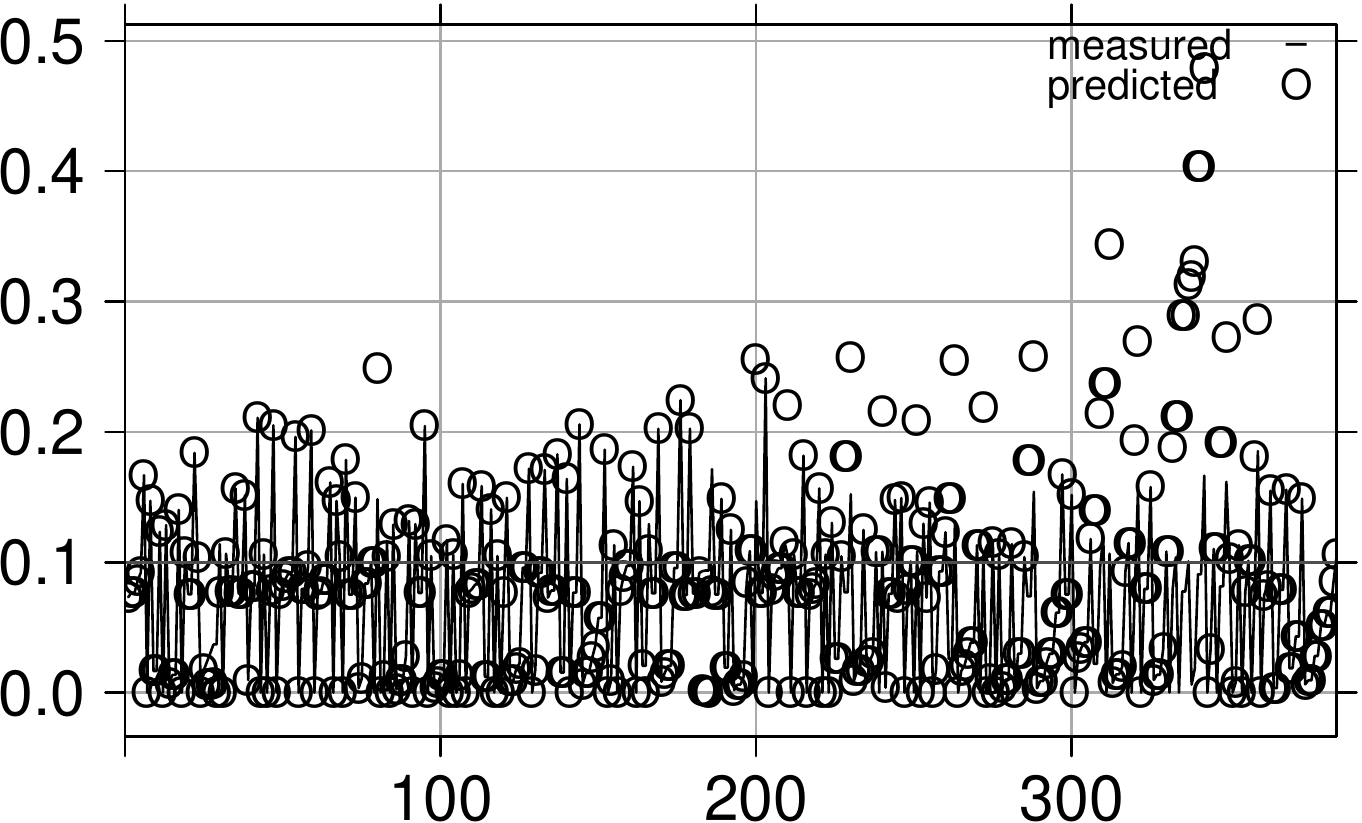}
    	\label{fig:error-h}
    }
    \subfigure[AQHI(20\%) Error] {    
    	\includegraphics[width=0.25\textwidth]{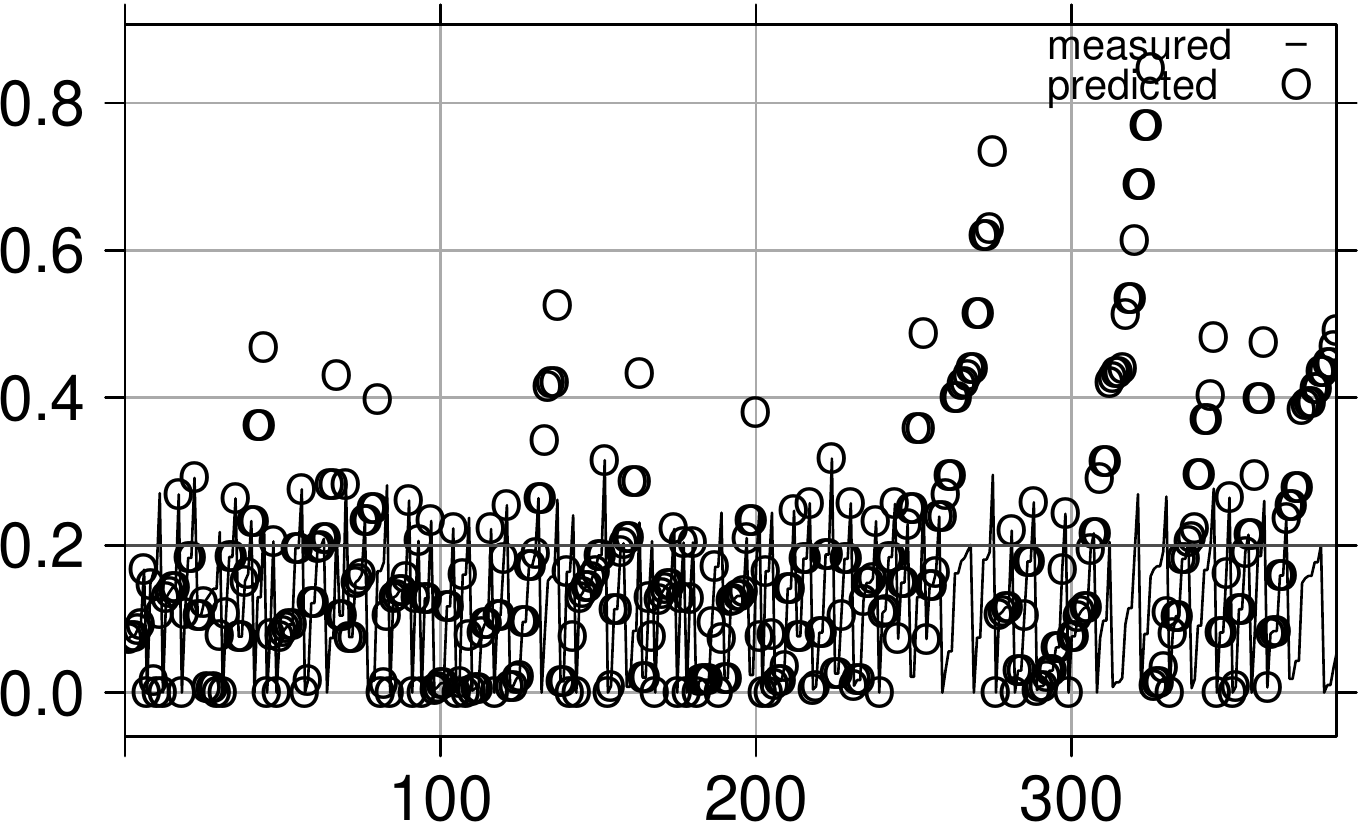}
    	\label{fig:error-i}
    } \\

    \subfigure[AQHI(5\%) Prediction Deviation] {    
    	\includegraphics[width=0.25\textwidth]{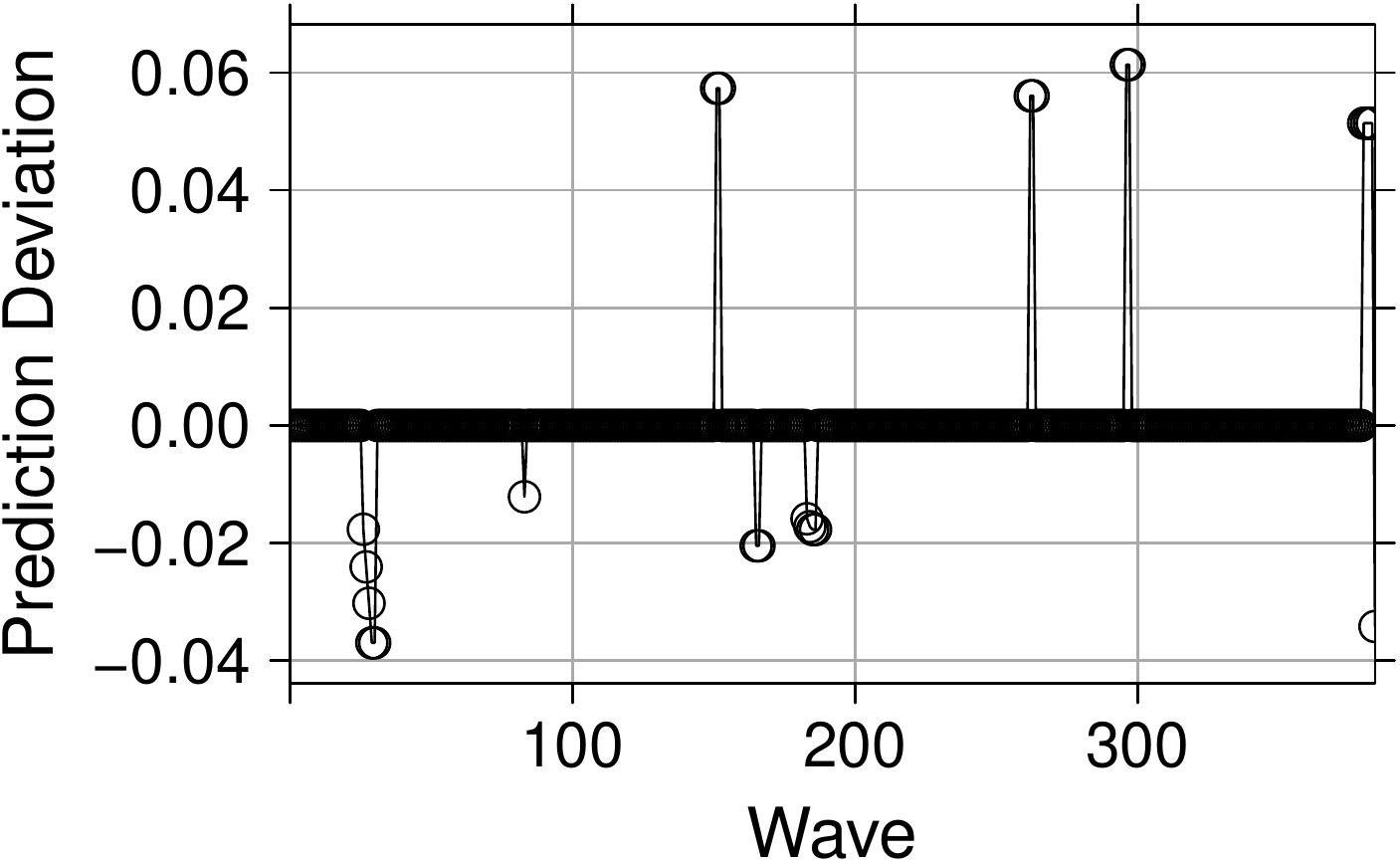}
    	\label{fig:error-j}
    }
    \subfigure[AQHI(10\%) Prediction Deviation] {    
    	\includegraphics[width=0.25\textwidth]{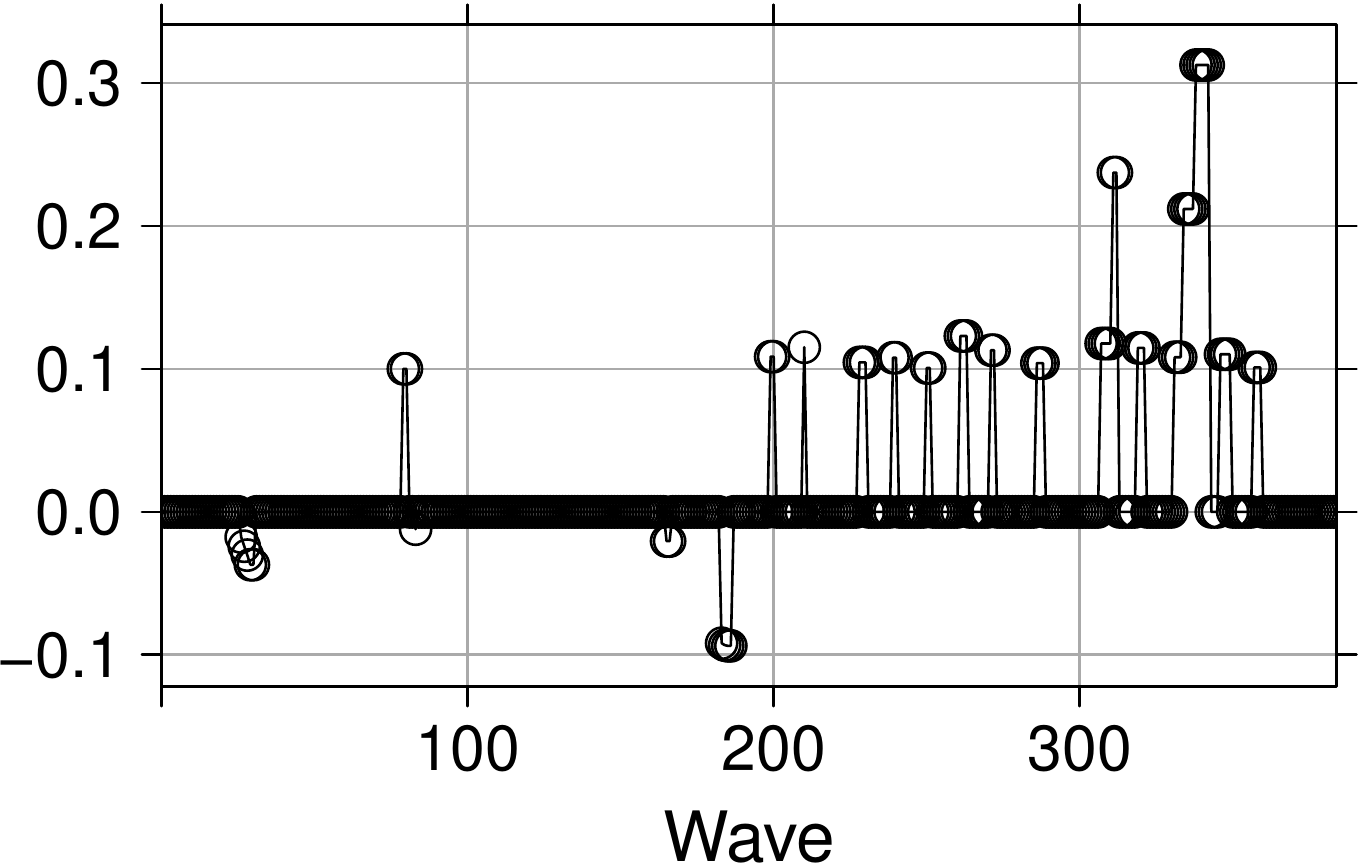}
    	\label{fig:error-k}
    }    
    \subfigure[AQHI(20\%) Prediction Deviation] {    
    	\includegraphics[width=0.25\textwidth]{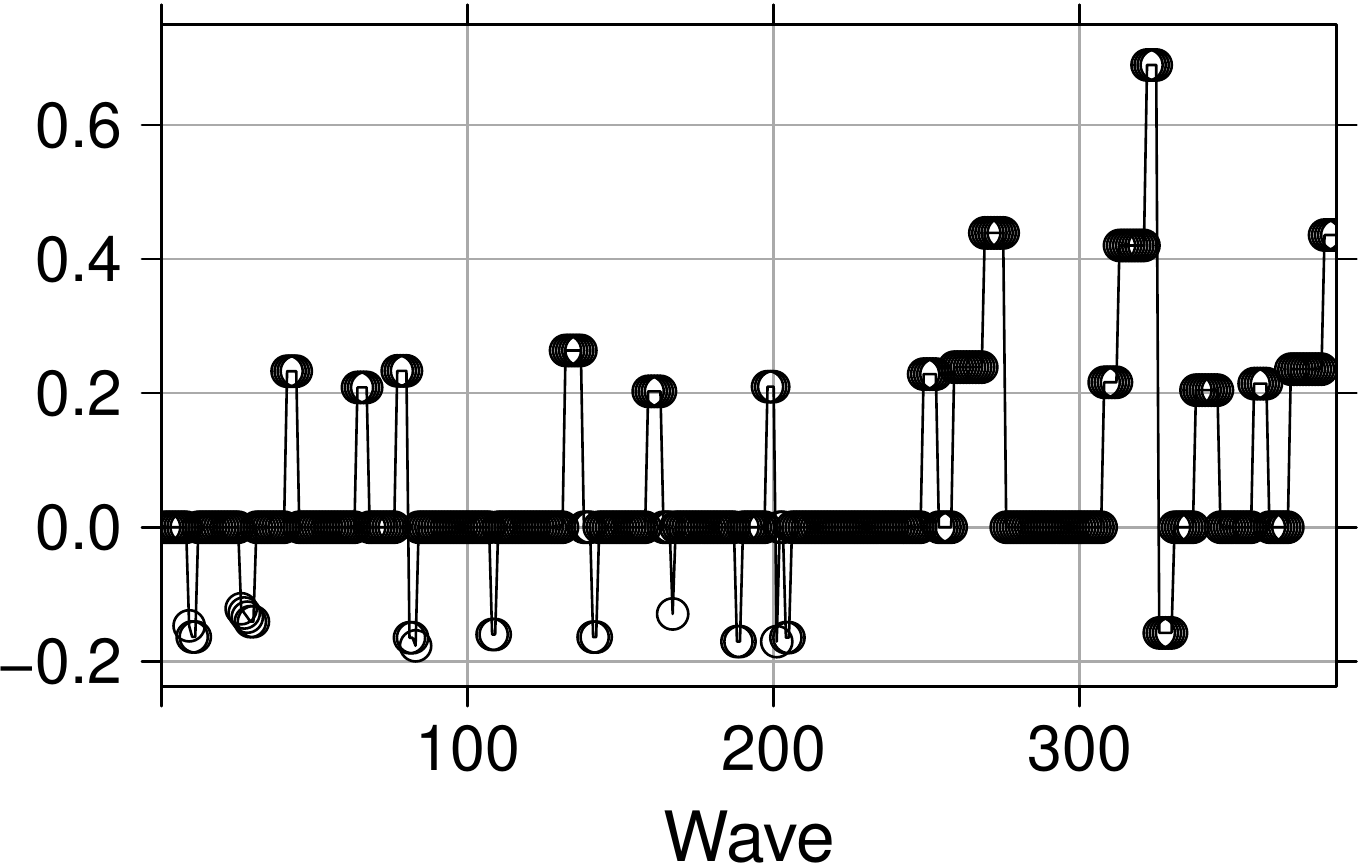}
    	\label{fig:error-l}
    } \\    
    
      
    \caption{Difference between measured and predicted error for the last processing steps of LRB and AQHI with error bounds of 5, 10, and 20\%}
    \label{fig:error}
    \vspace{-4mm}
\end{figure*}

Across waves, Figure~\ref{fig:error} shows the difference between predicted and measured errors for the last processing steps, that determine the workflow output, of LRB and AQHI using error bounds of 5,10, and 20\%. The predicted errors were calculated by accumulating the simulated errors (when compared against the output of synchronous executions), according to the binary values returned by the classifier across waves. Figures~\ref{fig:error-a}-\ref{fig:error-c}, \ref{fig:error-g}-\ref{fig:error-i}, show the predicted and measured errors in absolute value (Error); and figures~\ref{fig:error-d}-\ref{fig:error-f}, \ref{fig:error-j}-\ref{fig:error-l}, show the difference between predicted and measured errors (Prediction Deviation). A negative difference on a wave means that we were predicting the error below its actual (measured) value, and thus error bound violation did not happen for that wave. A positive difference on a wave means that the step was not executed and the predicted error stayed above $max_\varepsilon$. Globally, to maximize the ratio number-of-savings/number-of-violations, predicted and measured errors should be as close as possible, so that the prediction deviation goes to around zero most of the time, with the figures showing only markers for the outliers to the global trend.

For LRB with an error bound of 5 and 10\%, we can see that the predicted error stayed below the measured error for most of the time, with a deviation downto -0.1 (figures~\ref{fig:error-d}, \ref{fig:error-e}). When $max_\varepsilon$ was violated, 3 and 4 times with a bound of 5 and 10\% respectively, the difference between predicted $\varepsilon$ and $max_\varepsilon$ was never above 0.3, and only 1 time above 0.15 (subfigures~\ref{fig:error-d}, \ref{fig:error-e}). For a bound of 20\%, figures~\ref{fig:error-c}-\ref{fig:error-f}, the quality of the prediction was degraded: the predicted error exceeded $max_\varepsilon$ for a higher number of waves, albeit the prediction error was below 0.15 for most of the failed waves and below 0.45 for all waves. Nevertheless, $max_\varepsilon$ violation occurred in less than 10\% of the 500 waves, 82\% of which with minor violation ($<0.15$). Therefore, the potential for resource savings can be leveraged just at the expense of limited and mostly predictable additional error.

Regarding AQHI, subfigures~\ref{fig:error-g}-\ref{fig:error-l}, we can see that, with an error bound of 5\%, the deviation between predicted and measured error was minimal and $max_\varepsilon$ violation happened in only 4 waves ($<0.012$). With a bound of 10\%, more prediction errors arose after 200 waves, albeit never exceeding 0.32 overall and 0.10 for the majority. Finally, for a bound of 20\%, the number of prediction errors increases with errors staying below 0.6 overall and 0.25 for the majority. As a separate test, we optimized the classifier to maximize recall and obtained none prediction errors. However, many executions degraded resource efficiency, which lead us to keep the default parametrization.

To conclude, we may see that as larger is the error bound, the higher is the number and magnitude of the errors obtained on prediction. This is  expected as larger bounds on output difference allow for more (cumulative) deviation over time.

\myparagraph{Confidence Levels}
\begin{figure}
    \centering
	\subfigure[LRB] {
		\includegraphics[width=0.46\columnwidth]{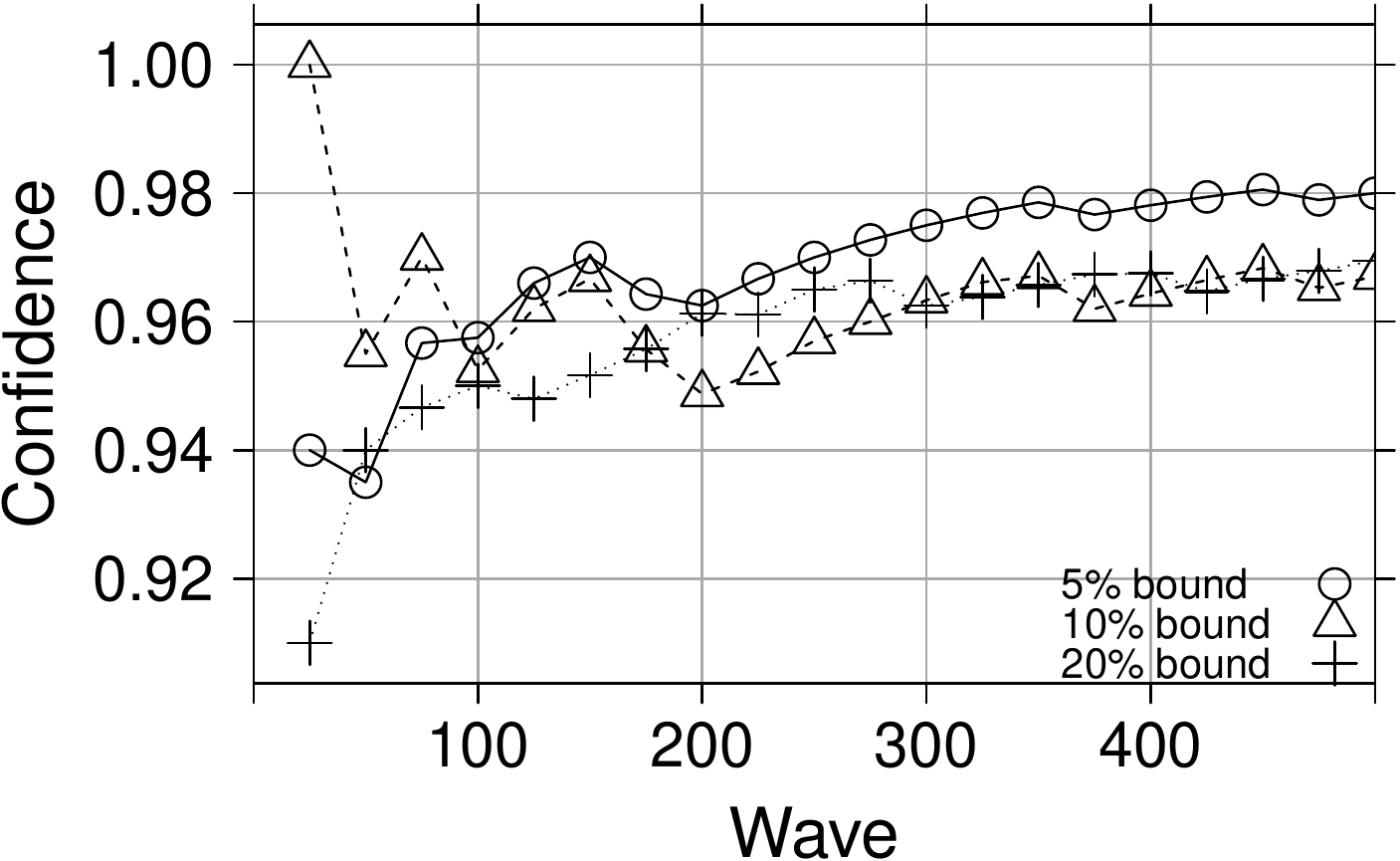}
	}
	\subfigure[AQHI] {
		\includegraphics[width=0.46\columnwidth]{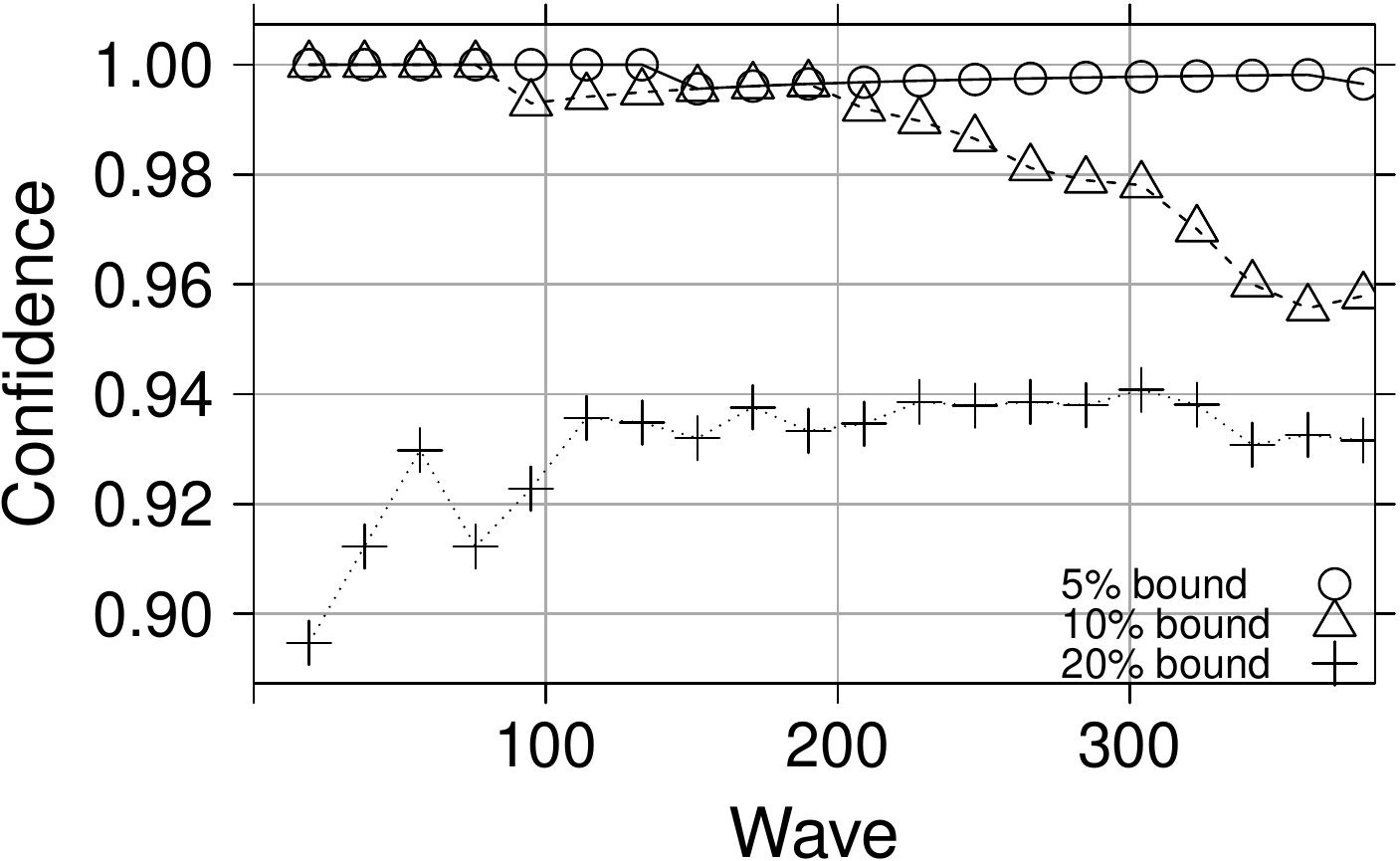}
	}
	\caption{Confidence in respecting error bounds as waves move forward}
	\label{fig:confidence}
\end{figure}%
\begin{figure}
	\centering
	\subfigure[LRB] {
		\includegraphics[width=0.46\columnwidth]{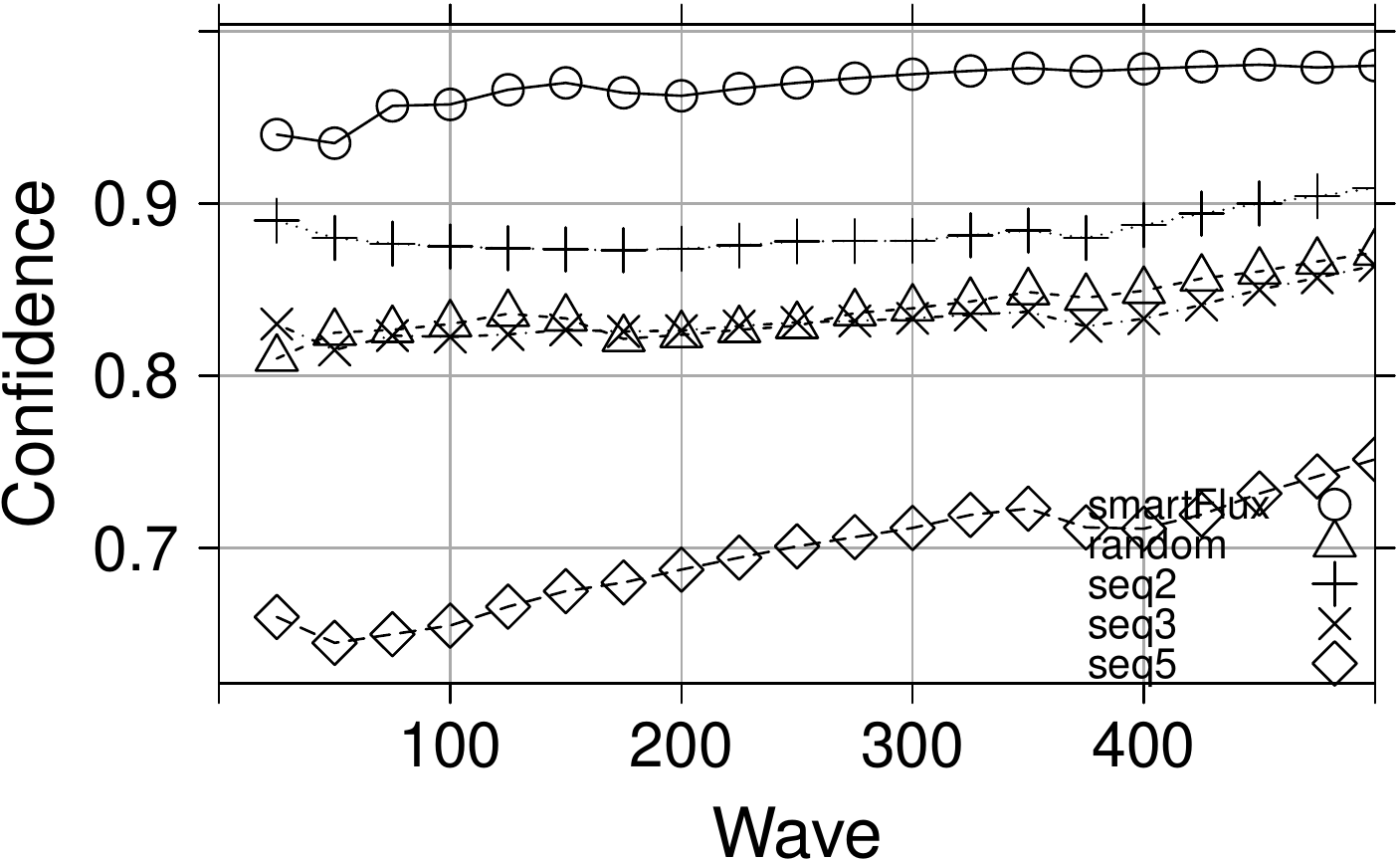}
	}
	\subfigure[AQHI] {
		\includegraphics[width=0.46\columnwidth]{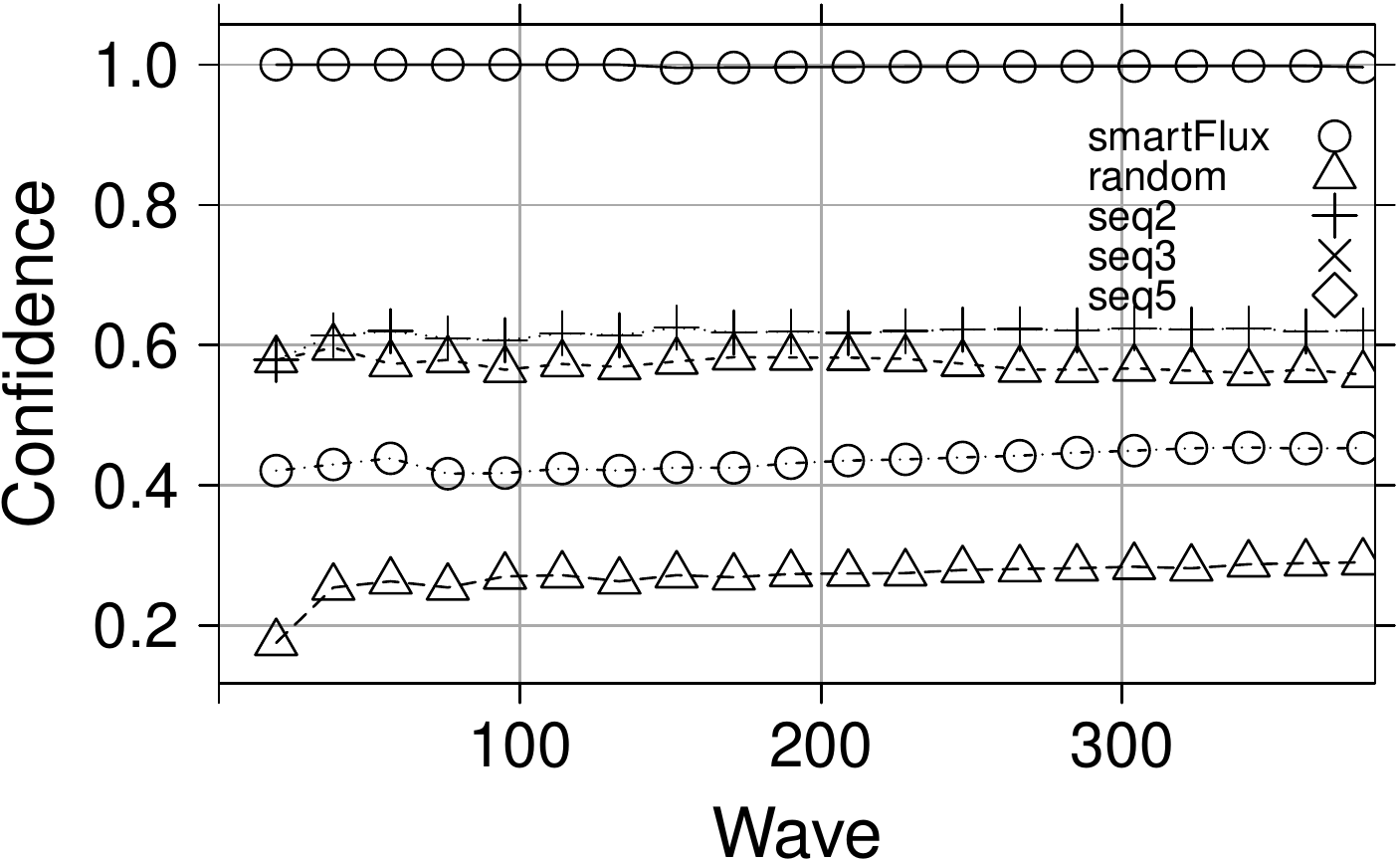}
	}
	\caption{Comparison of confidence levels for different triggering approaches with an error bound of 5\%}
	\label{fig:confidence-contrast}
\end{figure}
Figure~\ref{fig:confidence} shows, for LRB and AQHI, the confidence of our system in complying with defined error bounds, which corresponds to the normalized cumulative sum of correct waves where $max_\varepsilon$ was respected. We can see that, apart from the first 100 waves, the level of confidence was always above 95\% for error bounds of 5 and 10\% (i.e., for more than 95\% of the times we are able to comply with error bounds of 5 and 10\%). Nevertheless, with a bound of 20\%, the confidence level raised quickly to more than 95 and 90\% in LRB and AQHI respectively. This indicates that our system is reliable for decision makers. It can provide SLA-like guarantees stated as a confidence level (in \%, that can be regarded as a probability) of being (consistently) under a given error limit provided by the user. This is akin to current cloud SLAs that promise to honor availability (or limits to latency - a limit on time) for a given percentage of the time (that can also be regarded as a probability).


To show how well \name makes \emph{intelligent} decisions, we compare it with some naive approaches for an error bound of 5\% (this bound was selected in order to get the best possible confidence from these approaches). This comparison is given in Figure~\ref{fig:confidence-contrast}, where \emph{random} consists of randomly skipping step execution (executing or not executing a step on a given wave has equal probability), and \emph{seqX} consists of executing steps at every \emph{X} waves. We can observe that, either for LRB or AQHI, none approach was better than \name, which offers more than 95\% of confidence on error bound compliance. However, the other approaches revealed higher confidence in LRB than in AQHI, albeit never above 90\% for most part of the waves (note that a difference of 1\% in confidence is statistically significant). The reason to such difference in these workloads lies in the fact that LRB can be better approximated by a linear function than AQHI (\emph{seq2} has a pure linear behavior). However, only a Machine Learning approach can cover all cases (since polynomials can fit any type of correlation).

\myparagraph{Resource Savings}
\begin{figure*}
    \centering

    \subfigure[LRB Normalized Executions] {
    	\includegraphics[width=0.23\textwidth]{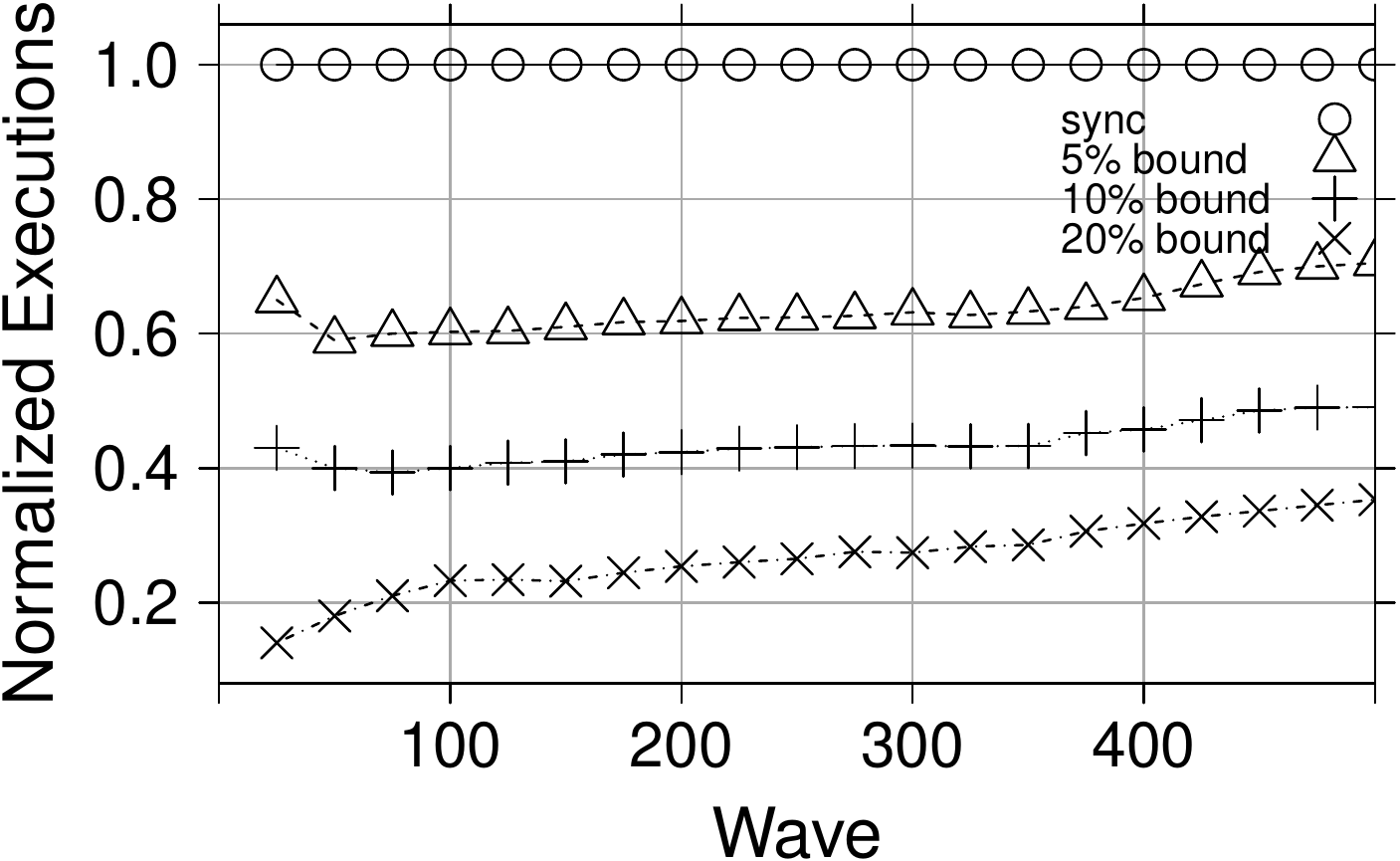}
    	\label{fig:resource-savings-a}
    }
	\subfigure[LRB Executions] {
		\includegraphics[width=0.23\textwidth]{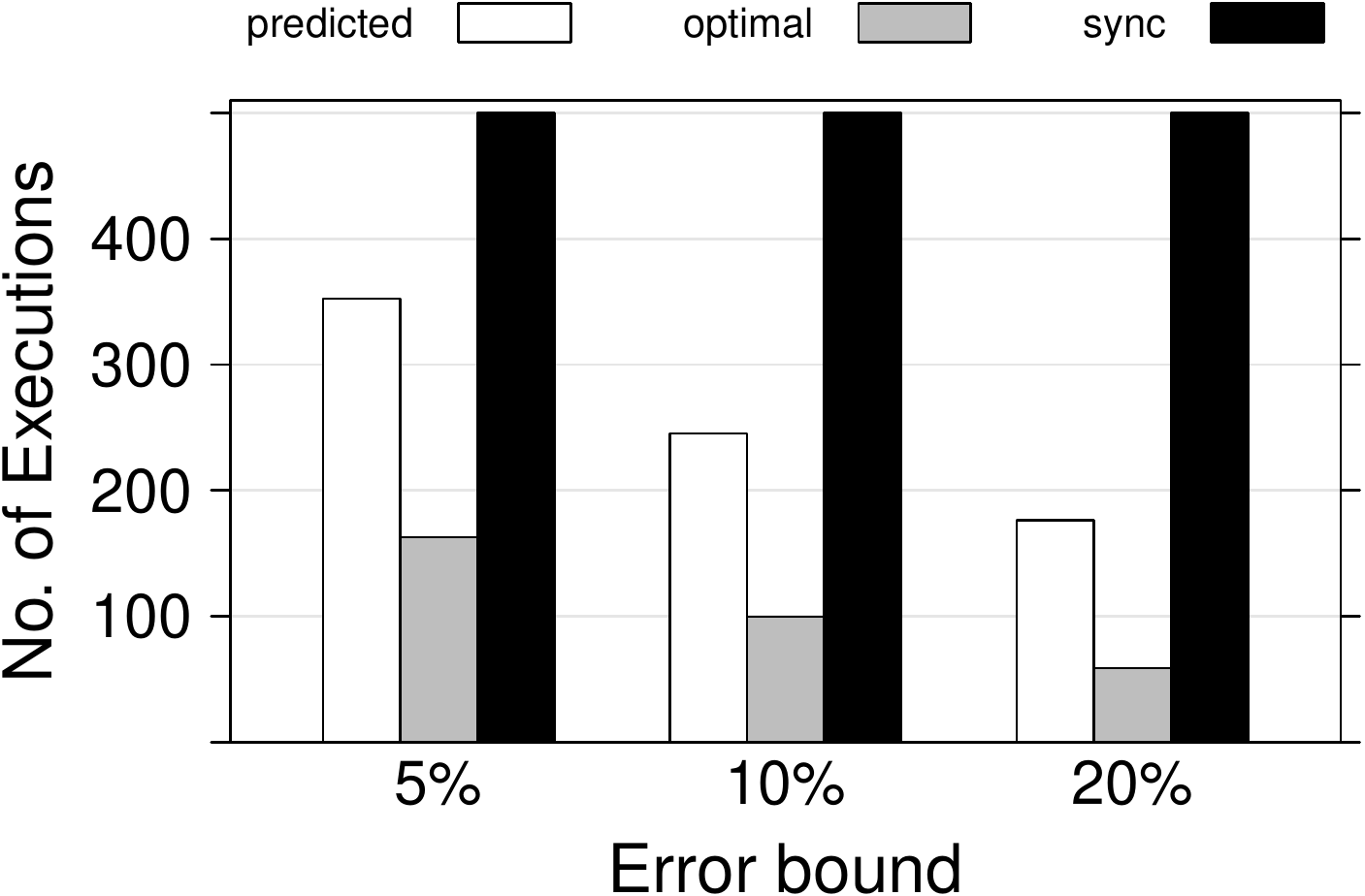}
    	\label{fig:resource-savings-b}		
	}
    \subfigure[AQHI Normalized Executions] {
    	\includegraphics[width=0.23\textwidth]{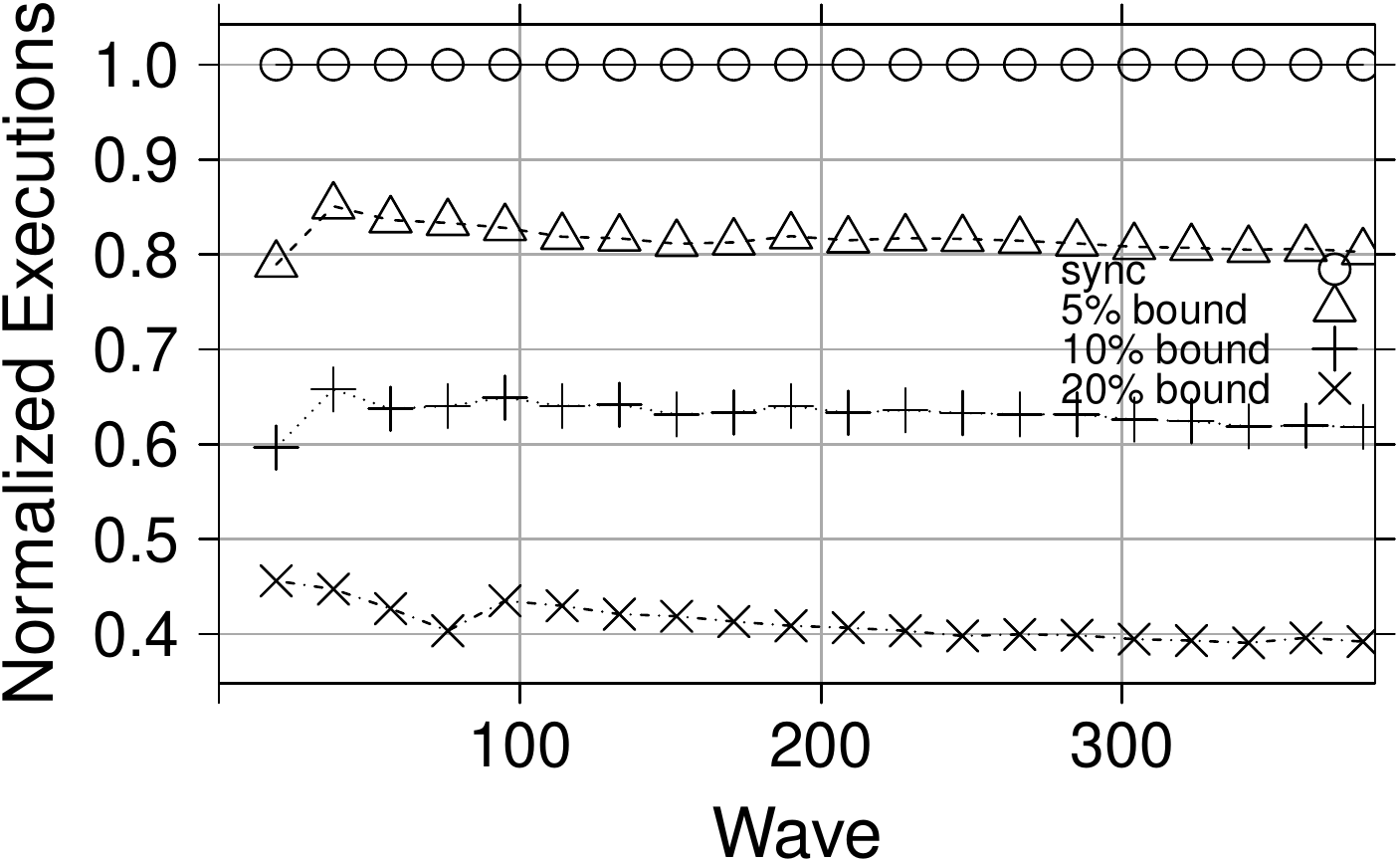}
    	\label{fig:resource-savings-c}
    }
	\subfigure[AQHI Executions] {
		\includegraphics[width=0.23\textwidth]{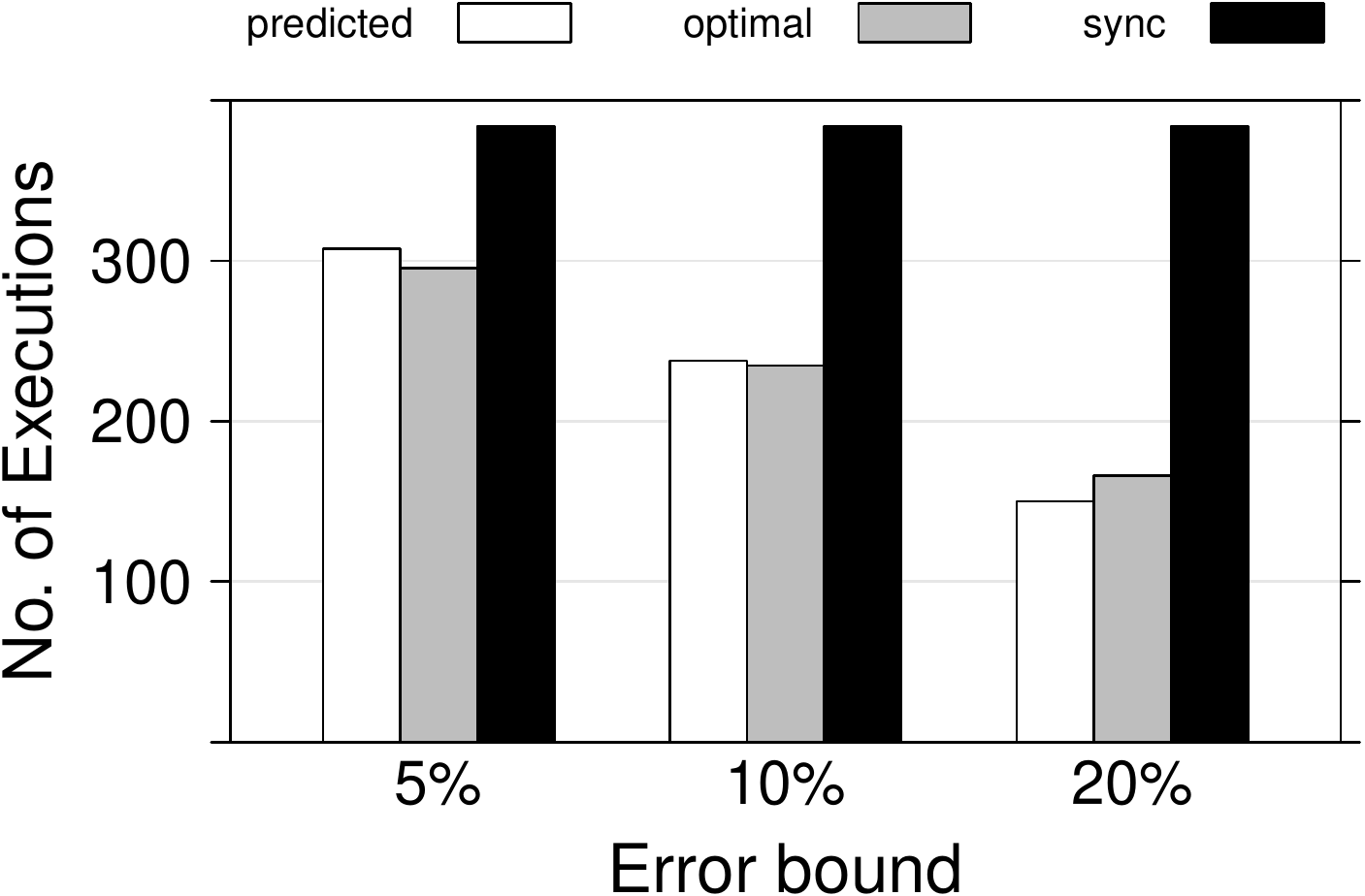}
    	\label{fig:resource-savings-d}
	}

    \caption{Executions performed with QoD versus synchronous model for LRB and AQHI}
    \label{fig:resource-savings}
\end{figure*}
Figure~\ref{fig:resource-savings} shows the executions performed and saved (resources engaged and spared) by \name against the regular synchronous model (SDF). For the cumulative sum of executions normalized over waves in LRB (Figure~\ref{fig:resource-savings-a}), we can see that, with only a bound of 5\%, the workflow steps were executed on average less than 70\% of the times in relation to SDF model; i.e., more than 30\% of the executions were saved, even for such a strict error bound. With a bound of 10 and 20\%, \name performed roughly 42 and 25\% of the executions respectively, leading to resource savings up to 75\%. Nonetheless, in Figure~\ref{fig:resource-savings-b} we may observe that we were not as efficient in saving executions as it would be optimal (i.e., delaying step triggering as much as possible without incurring in error violations, as it would be performed by a perfect fully-accurate predictor). This happened due to the optimization performed in the classifier to favor recall, leading to fewer saved executions yet to higher error compliance (which is usually more important for decision-making). If not, more than 50\% of the total predicted executions can be saved if the classifier is chosen to be more accurate (close to optimal).

For normalized executions in AQHI (Figure~\ref{fig:resource-savings-c}), we may see that the workflow is more stable, since the amount of saved executions is roughly the same across waves for each of the considered bounds. With a $max_\varepsilon$ of 5, 10 and 20\%, \name executes roughly 80, 60 and 40\% of the times respectively on average against the SDF model; hence, corresponding to 20, 40 and 60\% of saved executions as $max_\varepsilon$ increases. As the correlation between input impact and error was more uniform over time, the patterns of this workflow were better predicted, as shown in Figure~\ref{fig:resource-savings-d}: the total number of predicted executions was very close from the optimal number for each of the considered bounds.

With \name, we are thus able to save resources in exchange of allowing small but bounded errors to exist. As shown, roughly 20-30\% of unnecessary executions are saved for a bound of 5\%, and roughly 20-60\% are saved for bounds of 10\% and 20\%, which is substantial in a cloud environment, where resources are paid for or shared among a multitude of users and applications. This, while not degrading the quality of the applications and resulting information for decision makers. Further, we allow for user choices that achieve resource savings close to the optimal values that would never overrun the error bound.

\myparagraph{Overhead} There are the following sources of overhead: i) monitoring accesses to the data store ii) computing the input impact; iii) computing the output error; iv) writing the training set to disk; v) building the classification model; vi) persisting previous computation state; and vii) classifying instances with input impact values. We relied on Application Libraries to intercept read/write calls to the data store (cf. \ref{sect:architecture}) and on the equations presented in \ref{sect:model} to compute the input impact and output error. For each wave of data, we measured the running time of tasks that were executed with \name and compared with the time they take using the clean WMS version (without \name). The overhead for each task was always close to 0\%. Note that the overall overhead of the system, for a large bounded period, is negative, since we are skipping executions with \name. Building the classification model took the longest time (among all sources of overhead), albeit less than a second. Also, persisting previous state took roughly 0\% of overhead, since i) we set writings to HBase to be non-blocking; and ii) reads were part of requests to read the actual state (i.e., when retrieving column families from HBase, we get the column qualifiers corresponding to the actual and previous state in the same time as we were requesting only one column qualifier).

%
%
%
%

%

\section{Related Work}
\label{sect:related-work}

In the workflow domain, popular WMSs include: DAGMan, Pegasus, Taverna, Dryad, Kepler, Triana, Galaxy~\cite{WMS-survey-2015}. WMSs for the MapReduce (MR) Hadoop~\cite{White:2009:HDG:1717298}, like Oozie~\cite{islam:2012:ots:2443416.2443420}, also started to arise, e.g., Azkaban,\footurl{http://sna-projects.com/azkaban/} Cascading.\footurl{http://www.cascading.org}

More modern functionality in MR such as supporting social networks and data analytics are extremely cumbersome to code as a giant set of interdependent MR programs. Reusability is thus very limited. To amend this, DAG-like (workflow) platforms started to emerge on top of MR, such as Apache Tez\footurl{https://tez.apache.org/}
 and Pig \cite{Olston:2008:PLN:1376616.1376726}. The Apache Pig platform eases creation of data analysis programs. The Pig Latin language combines imperative-like script language (foreach, load, store) with SQL-like operators (group, filter). Scripts are compiled into Java programs linked to Map Reduce libraries.
The Hive \cite{Thusoo09hive-a} warehouse reinstates fully declarative SQL-like languages (HiveQL) over data in tables (stored as files in an HDFS directory). Queries are compiled into MR jobs to be executed on Hadoop. SCOPE \cite{Chaiken:2008:SEE:1454159.1454166} takes a similar approach to scripting but targeting Dryad \cite{Isard:2007:DDD:1272996.1273005} for its execution engine.

To avoid recreating web indexes from scratch after each web crawl, Google Percolator \cite{Peng:2010:LIP:1924943.1924961} performs incremental processing on top of BigTable, replacing batch processing of MR. It provides row and table-wide transactions, snapshot isolation, with locks stored in special Bigtable columns. Notify columns are set when rows are updated, with several threads scanning them. Applications are sets of custom-coded observers. Although it scales better than MR, it has 30-fold resource overhead over traditional RDBMS. Nova \cite{Olston:2011:NCP:1989323.1989439} is similar but has no latency goals, accumulating many new inputs and processing them lazily for throughput. Moreover, Nova provides data processing abstraction through Pig Latin; and supports stateful continuous processing of evolving data sets.

Yahoo CBP \cite{Logothetis:2010:SBP:1807128.1807138} aims at greater expressiveness by specifying incremental processing as dataflows with explicit mention when computation stages are stateless or stateful. Input is split by determining membership in frames of new records, allowing grouping input to reduce messaging.
CBP provides primitives for explicit control flow and synchronize execution of multiple inputs. It requires an extended MR implementation and some explicit programming to use a QoD-enabled dataflow.

Nectar \cite{Gunda:2010:NAM:1924943.1924949} for Dryad links data and the computation that generated it as unified hybrid cacheable element. On programs reruns, Nectar replaces results with cached data, which requires cache management calls that update the cache server. This is transparently done in InCoop \cite{Bhatotia:2011:IMI:2038916.2038923}, which does caching for MR applications. Map, combine and reduce phase results are stored and memoized.
Somehow like \name, this project attempts to reduce the number of executions; however, it implies that the input\slash output datasets are repeated or intersected among each other, whereas the QoD model fits a broader range of scenarios.

In \cite{asyncr_olston}, the authors present a formal
for defining temporal asynchrony in workflows.
The operators have signatures that describe the types and consistency of the blocks accepted as input and returned as output. Data channels have a representation of time to a relation snapshot, with an interval of validity, which are used to enforce consistency invariants. These constraints, types of blocks permitted on output, freshness and consistency bounds, are then used by the scheduler which produces minimal-cost execution plans. This project shares our goals of exploring and providing non ad-hoc solutions for introducing asynchronous behavior in workflows, however, it does not account with the volume, relevance or impact of modifications of the data given as input to each workflow step.

In~\cite{DBLP:journals/jisa/EstevesSV13}, which uses a mechanism inspired by~\cite{Esteves2012}, authors propose a DAG model where task triggering is based on 3 user-defined constraints: i) the time to trigger a task; ii) the number of updates on the data; and iii) the magnitude of the updates. This allows flexible data-based execution, however i) it is difficult to manually set a combination of constraints in a workflow in order to keep the error  within manageable levels;  and ii) no reasoning is performed about the impact computations have on varying the output. Thus, their model is not usable for scenarios where the freshness of the results needs to be guaranteed (like it is achievable with \name).

Further, it is important to note that we do not discard any data, like it happens in load shedding~\cite{Tatbul:2007:SFE:1325851.1325873}. In load shedding, a fraction of the input data is shed to alleviate overloaded servers and preserve low latency for query results. Contrarily to discarding data, we accumulate it up to the point where it causes significant changes on the output of the workflow. The observed errors happen not because we are making computations with incomplete data, but because we are not performing the computations and generating new output (i.e., errors come from stale data in the output).

There has also been a recent effort to enable approximate processing in data processing systems
in order to reduce latency (and possibly resource usage). However, these systems usually only target specific aggregation operators (e.g., sum, count) in structured languages~\cite{Goiri:2015:ABA:2694344.2694351,agarwal:2013:bqb:2465351.2465355,Agarwal:2014:KYW:2588555.2593667}. In our work, we provide approximate results for general-purpose computations; i.e., we are agnostic to the code that is running on each processing step and solely observe the data that is inputted and its effect on modifying the output. Effectively learning the correlation between input and output allows us to bound the error and give (probabilistic) guarantees about the correctness of the results. This makes \name unique.

\section{Conclusion}
\label{sect:conclusion}
We presented a novel workflow model, for continuous and data-intensive processing, capable of dynamically controlling the triggering of processing steps based on the predicted impact that input data might have on changing the workflow output. This impact, and level of triggering control provided, represents the QoD that governs the system to attain resource efficiency while maintaining results meaningful. To ensure correctness and freshness of these results, we bound the output deviation by making use of Machine Learning with Random Forests.
\\\indent We also proposed \name, a middleware framework implementing our workflow model that can be effortlessly integrated with existing WMSs. Experimental results indicate that we are able to save a significant amount of resources in exchange of allowing small bounded errors to exist (up to 30\% savings with a bound of 5\%). We provide compliance with error bounds with a high confidence level ($> 95\%$). The resulting savings can also be translated to less energy consumption, which is essential for a greener IT. \\\indent Overall, the results enable the creation of SLA-like guarantees on ensuring, effectively and efficiently, the quality of information provided to decision makers by workflow application results, clearly expressed as a (high) probability (a guarantee) of complying with a maximum defined tolerated error.


\section*{Acknowledgement}
We would like to thank Professor Andreas Wichert and his Ph.D graduate Catarina Moreira for their insight on Machine Learning aspects.

This work was supported by national funds through Funda\c{c}\~{a}o para a Ci\^{e}ncia e a Tecnologia with reference UID/CEC/50021/2013 and SFRH/BD/80099/2011.

\bibliographystyle{abbrv}
{\scriptsize\bibliography{refs-all}}


\end{document}